 \documentclass[prb,showpacs,preprintnumbers,amsmath,amssymb,preprint,12pt]{revtex4}

\usepackage[dvips]{graphicx}
\usepackage{color}
\usepackage{dcolumn}
\usepackage{bm}

\newcommand{\cbJ}{\mbox{\boldmath{$\cal J$}}}
\newcommand{\bmA}{{\mbox{\boldmath$A$}}}
\newcommand{\bmI}{{\mbox{\boldmath$I$}}}
\newcommand{\bmJ}{{\mbox{\boldmath$J$}}}
\newcommand{\bmx}{{\mbox{\boldmath$x$}}}
\newcommand{\bmy}{{\mbox{\boldmath$y$}}}
\newcommand{\bmv}{{\mbox{\boldmath$v$}}}
\newcommand{\bmw}{{\mbox{\boldmath$w$}}}
\newcommand{\PR}{{Phys. Rev. }}
\newcommand{\JPA}{{J. Phys. A: Math. Gen. }}

\begin{document}

\title{
Fat-tailed distribution derived from first eigenvector of symmetric random sparse matrix
}

\author{Hisanao Takahashi}

\affiliation{
The Institute of Statistical Mathematics, 10-3 Midori-cho, Tachikawa-shi, Tokyo 190-8562, Japan\\
}

\date{\today}

\begin{abstract}
Many solutions for scientific problems rely on finding the first (largest) eigenvalue and eigenvector of a particular matrix.  We explore the distribution of the first eigenvector of a symmetric random sparse matrix.  To analyze the properties of the first eigenvalue/vector, we employ a methodology based on the cavity method, a well-established technique in the statistical physics.

A symmetric random sparse matrix in this paper can be regarded as an adjacency matrix for a network.  We show that if a network is constructed by nodes that have two different types of degrees then the distribution of its eigenvector has fat tails such as the stable distribution ($\alpha < 2 $) under a certain condition; whereas if a network is constructed with nodes that have only one type of degree, the distribution of its first eigenvector becomes the Gaussian approximately.  The cavity method is used to clarify these results.
\end{abstract}


\maketitle

\section{Introduction}

Many problems in science can be reduced to the eigenvalue/vector problem. The first eigenvalue/vector have sometimes particularly important meanings in these problems. For example, the first eigenvector of the transition probability matrix in physics represents the largest transfer direction.  In quantum physics, the assessment of the ground state is generally formulated as a first eigenvalue/vector problem\cite{Quantum}.  To analyze the spin-glass system, we often use the spin-glass susceptibility as an indicator of the critical phenomena. The spin-glass susceptibility is derived from the covariance matrix of its spins, so that the first eigenvalue of the correlation matrix plays an important role, especially at the critical point\cite{Takahashi,Aspelmeier}.

We show the density function of the first eigenvalues/vectors that are evaluated from the adjacency matrices of the networks whose nodes have two different types with respect to their degrees in this paper.  The adjacency matrix of this network is also reduced to a subset of the random sparse matrices and it seems that the property of the density function of the eigenvalues/vectors on such matrices has not been clarified very well so far, especially on the first eigenvector distribution.

We show the dependence of the density function on the ratio of two different degrees in this paper.  If a network is constructed with two different types of nodes with respect to the degree then the distribution of its first eigenvector has fat tails such as the stable distribution ($ \alpha < 2 $, $ \beta \simeq 0 $) under a certain condition; whereas if a network is constructed with only one type of degree, the distribution of its first eigenvector is the Gaussian distribution, approximately. 

We calculate the first eigenvalue/vector that is derived from taking the large system limit, using our developed scheme\cite{Kabashima,Kabashima+t} based on the cavity method and the scheme is applicable to a wide variety of networks.  In the population dynamics method which we use to assess the density function of the cavity field of the network, we need to employ the sequential update strategy for a stable convergence, which will be explained later.  
We evaluate two dimensional density function of the cavity fields in this paper, whereas we mainly discussed the case that the independence can be assumed for two variables of the cavity fields in Ref.~[\onlinecite{Kabashima}], i.e. $ q(A,\, H) \simeq q(A)\,q(H) $.

We also explore a Poissonian network and a Laplacian matrix for comparison. 
We show that the Poissonian network and the network whose nodes have two different types of degrees with a certain ratio have many similar properties.  This similarity is clarified by the results of the cavity method.  This means that many network properties are defined by the ratio of nodes that have the largest degree.  We show the shape of the density functions, whereas we used the inverse participation ratio (IPR) to evaluate the skewness of the density function of the eigenvectors in Ref.[\onlinecite{Kabashima+t}].  We found a richer structure in the shape of the density function and we will describe those in this paper.
The Poisonian netwrok is one of the most well-known Erd\"{o}s-R\'{e}nyi model which have been widely studied in network science\cite{ErdosRenyi}.
The Poissonian network is also related with the graph bisection problem\cite{CojaOghlan}.

Using the diagonal matrix and the adjacency matrix which can be regarded as a random symmetric matrix, we can compose a Laplacian matrix.  
Random impedance networks can be represented by a Laplacian matrix and we can evaluate the density of resonances of the networks as the spectral density of the eigenvalues\cite{Lapl1,Lapl2}.
The well-known Google PageRank$^{\rm TM}$ ranks World Wide Web pages on the basis of the first eigenvector of the Laplcian matrix whose entries represent the number of links of a huge network constructed with Web pages\cite{PageRank}.  
In networks science, the first eigenvector of the adjacency matrix is called the eigenvector centrality.

Many properties have been clarified for the spectral density of the eigenvalues that are evaluated from the ensembles of large random matrices.  We can find this clarified knowledge in random matrix theory and its related topics\cite{RandomMatrix, spar1}.
In random matrix theory, we also find a knowledge on the {\it first} eigenvalue distribution of dense matrices which is called as the Tracy-Widom distribution.
However, relatively little is known about the distribution of the eigenvalues/vector for the random sparse matrices\cite{Sodin,BrayRodgers,BiroliMonasson,Semerjan,Kuhn,Metz,Takeda}.
The problem which can be described by a random matrix with a constraint is related to a wide region of science, e.g., combinatorial problems in computer science\cite{Alon}, statistical properties of disordered conductors and of chaotic quantum systems\cite{Guhr}, since Wigner designed the random matrix theory to deal with the statistics of eigenvalues/vectors of complex many-body quantum systems\cite{Wigner}.

This paper is organized as follows. 
The next section introduces the model that will be explored. 
In section \ref{Cavity approach for}, we describe a scheme for examining the eigenvalue/vector problem in a large system limit on based on the cavity method. 
In section \ref{Calculation methodology}, we discuss some numerical methods used in this paper.  
The results are described in section \ref{Results}.
Concluding remarks are presented in the final section.

\section{Model definition}
In this section, we describe the method of constructing the network which is discussed in this paper.  For this purpose, we make use of an adjacency matrix which uniquely defines the network structure.  We construct the network whose nodes have two different types of degrees and it will be referred as 2-DTD network or 2-DTD model hereafter.  This network can be regarded as one of the simplest cases of a network that generally has various types of nodes with respect to the degree.  
We also explore the networks whose degree distribution is the Poissonian distribution and that might be one typical case among networks that has various types of nodes with respect to the degree.  This network is often mentioned as a model for the World Wide Web network and other scientific structures.  We also find it in the bisection problem.

\subsection{Network contraction algorithm and adjacency matrix}
Consider a $N \times N$ real symmetric sparse matrix $\bmJ=(J_{ij})$ that is characterized by a distribution $p(k)$, where $k(=0,1,2,\ldots)$ denotes the number of non-zero entries in a column/row of the matrix and represents the degree of the corresponding node.  We set the diagonal elements of the matrices to zero, because we assume that there are no self-loops in the network.  
We define $d_i$ that represents the degree of a node indexed $i(=1,2,\ldots,N$) as follows: We draw a number $ k $ from the stochastic variable obeying the distribution $p(k)$ and set $d_1=k$, and repeat the same procedure for all $ i$.

Now, we randomly decide the non-zero entries in a $N \times N$ adjacency matrix which represents the links of the network as the following algorithm\cite{StegerWormald}:

\begin{itemize} 
\item[(S)] Prepare a set of indices $U$ in that each index $i$ attends $d_i$ times.  
\item[(A)] Repeat the following until no suitable pair can be found. Choose a pair of elements $(I,\, j)$ from $U$, randomly.  If $i \ne j$ and the pair $(I,\, j)$ has not been chosen before, then make a link between them and remove the two elements $i$ and $j$ from $U$. Otherwise, we return them back to $U$. 
\item[(B)] If $U$ is empty, finish the algorithm.  Otherwise, return to (S) and start over again. 
\end{itemize}
We set 1 to the elements of the adjacency matrix as corresponding with the link of the above network.  This construction agrees with the usual definition of adjacency matrix. 

Here, we modify the above adjacency matrix as follows.  We replace the values of the non-zero entries symmetrically i.e. $J_{ij}=J_{ji}$ , and stochastically obeying the following probability:
\begin{eqnarray} 
p_{\rm J}(J_{ij}|\Delta)= \frac{1+\Delta}{2}\delta(J_{ij}-1) + \frac{1-\Delta}{2} \delta(J_{ij}+1),
\label{binary_dist} 
\end{eqnarray}
where $\delta(x)$ denotes the Dirac delta function and $0\le \Delta \le 1$.  The $ \Delta$ controls the number of the negative entries which represent the links between two different groups in the bisection problem.  We interpret the negative entries as frustrations that are often discussed in the spin glass model.

In this paper, we discuss a property of an ensemble average of the modified adjacency matrices that generate from the same algorithm.  For this purpose, we made the thousands of networks, running the above algorithm.

\subsection{2-DTD network and Poissonian network}
Within the set of networks that have two different types of nodes with respect to the degree, we mainly focus on the networks whose nodes are the 4 or 8 degrees with ratio 0.9:0.1:  
\begin{eqnarray}
	p(k)= \left\{
		\begin{array}{ll}
			0.9 & \mbox{if $ k=4$,}\\
			0.1 & \mbox{if $ k=8$,}\\
			0   & \mbox{otherwise.}
		\end{array}
	\right.
\label{pktwod}
\end{eqnarray}
For comparison, we also show results on several other cases as follows: the ratio is different from the above case, the value of the larger degree is different, and the network that all nodes are the same with respect to the degree.

\begin{figure}
\rotatebox{90}{\hspace{25mm}\rotatebox{0}{density}}\hspace{-5mm}
\includegraphics[width=6cm]{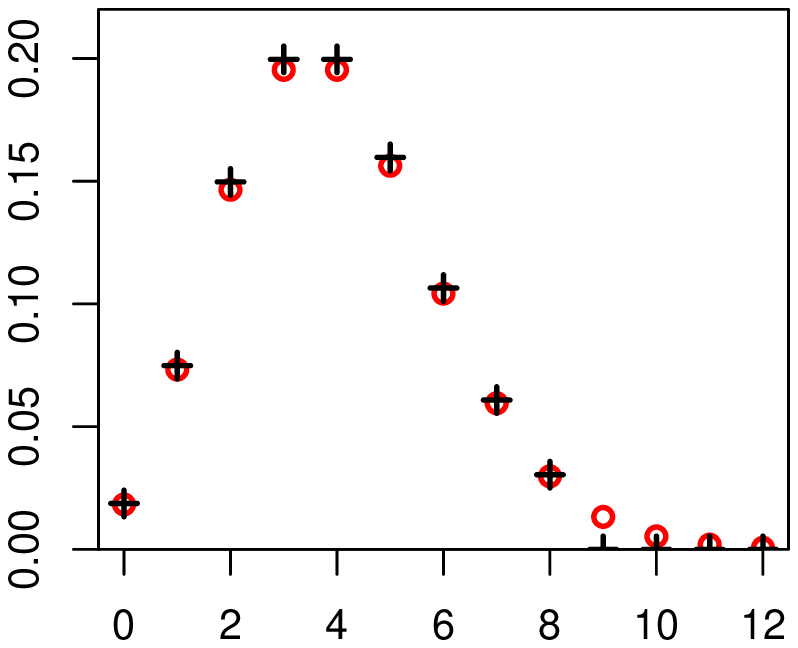}\vspace{-12mm}\\
\hspace{5mm} $ k $
\caption{Modified Poisson density $p(k) $ and the ordinal Poisson density $ p^{\rm poi}(k) $. Crosses represent the modified Poisson density defined by Eq.~(\ref{normalize-poissonian}) and circles represent the ordinal Poisson density.}
\label{fig-Poissondens}
\end{figure}

As regards a Poissonian network, if the support of the number of degrees is not bound by a maximum value, i.e. the number of non-zero entries in each row/column is infinite, the first eigenvalue generally diverges as $N \to \infty$.
To avoid this phenomenon, we assume that $p(k)=0$ for $k$ larger than a certain value which is denoted by $k_{\rm max}$.  To normalize the probability distribution, we modify the degree distribution for the Poissonian network as
\begin{eqnarray}
\label{normalize-poissonian} 
	p(k)=\frac{1}{p_{\rm nor}}p^{\rm poi}(k), \qquad 
	\frac{1}{p_{\rm nor}}\sum_{l=0}^{k_{\rm max}} p^{\rm poi}(k) = 1,
\end{eqnarray}
where $ p^{\rm poi}(k)=\lambda^k\exp(-\lambda)/k! $ is the original Poisson distribution and $ p_{\rm nor} $ is a normalization factor.
In this paper, we explore the case shown in Fig.\ref{fig-Poissondens} in which is $k_{\rm max}=8$ and $ \lambda=4 $. The results of the Poissonian Network have many similarities with the results of 2-DTD model of Eq.~(\ref{pktwod}) as will be shown later.

\section{Cavity approach for first eigenvalue problem}
\label{Cavity approach for}
The cavity method, a well-known method in physics, has be applied to many problems\cite{Kuhn,Metz,Takeda,Cavity,Kabashima,Kabashima+t}.  In this paper, we use the two-dimensional density function of the cavity fields.  Using this method with the sequential update strategy, we obtain stable results for a wider variety of networks including 2-DTD network and the Poissonian network.  To obtain the stable results in the case that the density function of the eigenvector has a fat tail, we need to modify our previous method in Ref.~\onlinecite{Kabashima}.  In order to verify our results, we compare the results obtained by the cavity method with those evaluated by the power method.

\subsection{First eigenvalue/vector and message passing algorithm}
The first eigenvalue/vector problem can be formulated as an optimization problem:
\begin{eqnarray}
\min_{\bmw} \left \{-\bmw^{\rm T}\bmJ \bmw \right \} \ {\rm subject \ to} \ |\bmw|^2 = N,
\label{constrained}
\end{eqnarray}
where $ \displaystyle\min_{\bmw}\{\cdots \}$ denotes the minimization with respect to $\bmw$.  From the above optimization, we derive the optimal values of $\bmw$ which equal to the eigenvector $\bmv$ of the matrix $ \bmJ$.  The first eigenvalue $\Lambda$ is evaluated as $\Lambda=(\bmv)^{\rm T} \bmJ \bmv/N$.

When $N\to \infty$, if the distribution of the $v_i$ is the Cauchy distribution, then the variance is infinity.  In this case, the following relationship holds:
\[
 \frac{1}{N^{3/2}}\sum_i v_i^2 \sim O(1).
\]
Therefore, we need to modify the formulation of our problem (\ref{constrained}) as
\begin{eqnarray}
\mathop{\rm min}_{\bmw} \left \{-\bmw^{\rm T}\bmJ \bmw \right \} \ {\rm subject \ to} \ |\bmw|^2 = N^\xi,
\label{optimizationNxi}
\end{eqnarray}
Fortunately, the methodology we will explain is applicable in this case and Eqs.~(\ref{cavity_dist_update}) and (\ref{fulldist}) are hold, although we may need a small modification for the convergence test.  We will only explain the case of $\xi=1$. It is, however, easy to accommodate the description hereunder to other values of $\xi$.

The optimization (\ref{constrained}) can be performed by the method of Lagrange multipliers.  The Lagrange function is
\begin{eqnarray}
{\cal L}(\bmw,\lambda) &=& -\bmw^{\rm T} \bmJ \bmw +\lambda (|\bmw|^2-N) \cr &=& \lambda \sum_{i=1} w_i^2 - 2 \sum_{i > j} J_{ij}\, w_i\, w_j - \lambda N,
\label{Lagrangefun}
\end{eqnarray}
where $\lambda$ is a Lagrange multiplier.

Focusing on the site indexed $i$ and its surroundings, Eq (\ref{Lagrangefun}) is decomposed into
\begin{eqnarray} 
{\cal L}_i(w_i|A_i,H_i)=A_i\, w_i^2 - 2 H_i\, w_i
\label{single_quadratic_function} 
\end{eqnarray}
where the coefficients of the second and first order terms, $A_i$ and $H_i$, which are called as the cavity field, are determined in a certain self-consistent manner, i.e., by the cavity method.

\begin{figure}
\vspace{13mm}
\hspace{7mm}$ l $\vspace{-13mm}\\
\includegraphics[width=7cm]{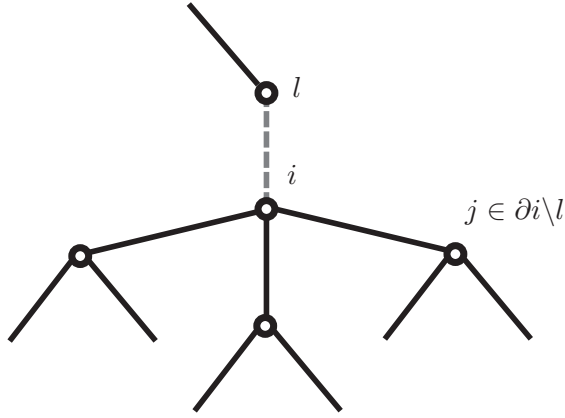}\vspace{0mm}\\
\vspace{-38mm}
\hspace{7mm}$ i $ \vspace{-3mm}\\
\hspace{65mm}$ j \in \partial i \backslash l $\vspace{23mm}\\

\caption{Network that can be assumed locally a tree.}
\label{Tree}
\end{figure}

To find the self-consistent values of $A_i$ and $H_i$, we introduce auxiliary variables $A_{j \to i}$ and $H_{j \to i}$, which respectively represent the second and first order coefficients of $j \in \partial i$, where the notation $ \partial i$ is the set of nodes connecting directly with node $i$, see Fig.~\ref{Tree}

If we regard our graph as locally a tree, then we can describe the local Lagrange function which is considered only the descendant of node $i$,
\begin{eqnarray}
{\cal L}_{i, \partial i \backslash l}(w_i,\{w_{j \in \partial i \backslash l}\}) =\lambda\, w_i^2 -2\, w_i \sum_{j \in \partial i \backslash l}J_{ij}\, w_j + \sum_{j \in \partial i \backslash l} \left (A_{j \to i}\, w_j^2 - 2\, H_{j \to i}\, w_j \right ). 
\label{localLagrangefunc}
\end{eqnarray}
To minimize the above function, we partially differentiate with respect to $w_j$. Then we obtain the relation,
\begin{eqnarray}
w_j=\frac{v_i\, J_{ij}+H_{j \to i}}{A_{j \to i}}. 
\label{plocalLagrangefunc}
\end{eqnarray}
Substituting the above relation into the function (\ref{localLagrangefunc}), and if we compare with the function 
$ {\cal L}_{i\to l} = A_{i \to l}\, w_i^2 - 2\, H_{i \to l}\, w_i  $,
then we find the following relationships:
\begin{eqnarray}
&&A_{i \to l} = \lambda- \sum_{j \in \partial i \backslash l} \frac{J_{ij}^2}{A_{j \to i}},
\label{cavity2} 
\\
 && H_{i \to l} = \sum_{j \in \partial i \backslash l} \frac{J_{ij} H_{j \to i}}{A_{j \to i}}. 
\label{cavity1}
\end{eqnarray}

Under a given initial condition, we evaluate the above equations and calculate the following values:
\begin{eqnarray}
	A_i
	&=&
	\lambda - \sum_{j \in \partial i} {J_{ij}^2}/{A_{j \to i}}
	\\
	H_i
	&=&
	\sum_{j \in \partial i} J_{ij} H_{j \to i}/A_{j \to i}
\label{cavity4}
\end{eqnarray}
If we use the right value of $\lambda$, i.e., the first eigenvalue of the matrix $ \bmJ$ then $ A_i $ and $ H_i $ become identical except for a numerical error even if the operation of Eqs. (\ref{cavity2})-(\ref{cavity4}) are performed again.  The above procedure offers the exact result when the graph is free from cycles.  However, when the graph contains cycles, the above algorithm is available to obtain the approximate results.  The cycles lengths in the connectivity graph, constructed by random sparse matrices, typically grow as $O(\ln N)$ when $N \to \infty$ \cite{RandomGraph}, and thus, we can ignore the effects of the cycles.

\subsection{Cavity fields and population dynamical method}
We apply a macroscopic approximation to the cavity fields and describe those as a two-dimensional distribution, i.e.,
\begin{eqnarray*}
q(A,H)\simeq\left(\sum_{i=1}^N n_i\right)^{-1} \sum_{i=1}^N \sum_{j \in \partial i} \delta(A-A_{j \to i})\, \delta(H-H_{j \to i}), 
\end{eqnarray*}
where $ n_i $ is the number of nodes directly connecting to node $i$.

When we choose an edge randomly and observe one terminal of the edge, the probability that the degree of the node is $k$ is described as
\begin{eqnarray}
r(k)=\frac{k\, p(k)}{\sum_{k=0}^{k_{\rm max}} k \, p(k)}.
\label{posterior}
\end{eqnarray}
Using this probability, we can describe the following self-consistent equation of $q(A,H)$, which is consistent with eqs. (\ref{cavity2}) and (\ref{cavity1}).
\begin{eqnarray}
&&
q(A,H) 
\nonumber
\\
\nonumber
&=&
\sum_{k=1}^{k_{\rm max}} r(k) \int \prod_{j=1}^{k-1} dA_j\, dH_j\, q(A_j,H_j)
\\
&&\qquad
\left \langle \delta\left (A-\lambda+\sum_{j=1}^{k-1}\frac{{\cal J}_j^2}{A_j} \right ) \delta\left (H-\sum_{j=1}^{k-1} \frac{{\cal J}_j H_j}{A_j} \right ) \right \rangle_{\cbJ} \! ,
\label{cavity_dist_update}
\end{eqnarray}
where $\left \langle \cdots \right \rangle_{\cbJ}$ represents the operation that takes the average with respect to ${\cal J}_j$ and ${\cal J}_j$ obeys $p_{\rm J}({\cal J}_j)$. After $q(A,H)$ is determined from this equation, the distribution of the auxiliary variables in the original system, i.e.,
\begin{eqnarray*}
Q(A,H)\simeq N^{-1} \sum_{i=1}^N \delta(A-A_i)\delta(H-H_i),
\end{eqnarray*}
is evaluated as 
\begin{eqnarray}
&&
Q(A,H)
\nonumber
\\\nonumber
& = &
 \sum_{k=0}^{k_{\rm max}} p(k) \int  \prod_{j=1}^k  dA_j\, dH_j\, q(A_j,H_j)
\\
&&\qquad
\left \langle \delta\left (A-\lambda+\sum_{j=1}^{k}\frac{{\cal J}_j^2}{A_j} \right ) \delta\left (H-\sum_{j=1}^{k} \frac{{\cal J}_j H_j}{A_j} \right ) \right \rangle_{\cbJ} \!.
\label{fulldist}
\end{eqnarray}
The population dynamical method was used to evaluate Eqs.~(\ref{cavity_dist_update}) and (\ref{fulldist}).  The densities $ q(A,H) $ at the right and left of Eq. (\ref{cavity_dist_update}) became identical only when we substitute the appropriate eigenvalue $\lambda$ and density $ q(A,H) $ in the right side of the equation.  In other cases, the left density function is different from with right one.  

To find the appropriate value for $\lambda$ which must equal to the first eigenvalue $\Lambda$, we evaluate the following statistics for several trial values of $\lambda$:
\begin{eqnarray*}
T =\int dA\, dH\, Q(A,H) (H/A)^2 \ (\simeq N^{-1}|\bmv|^2).
\end{eqnarray*}
On the basis of pairs of $\lambda$ and $ T $ from the above results, we estimate the value of $\lambda$ when $T$ equals to 1,  This method is not easy to use when the density has a fat tail such as in the stable distribution ($ 1 < \alpha<2$), because the density decays as $ x^{-(1+\alpha)}$, i.e., $T \to \infty$.  In that case, we need to employ another statistic such as the average of the absolute value of $v_i $, i.e.,
\begin{eqnarray*}
U=N^{-1}\sum_i |\bmv_i| =\int dA\, dH\, Q(A,H) |(H/A)|.
\end{eqnarray*}

In the population dynamical method, the distribution $q(A,H)$ is decomposed into many pairs of $A$ and $H$, i.e. $\{(A_1^{(t)},H_1^{(t)}),\, (A_2^{(t)},H_2^{(t)}),...,(A_n^{(t)},H_n^{(t)})\} $.  Using this set in the right side of Eq.~(\ref{cavity_dist_update}), we get the set of $A$ and $H$ for the next step $ t+1 $, $\{(A_1^{(t+1)},H_1^{(t+1)}),\, (A_2^{(t+1)},H_2^{(t+1)}),...,(A_n^{(t+1)},H_n^{(t+1)})\} $.  To continue this procedure, there exit many update strategies. Here, we used the sequential update strategy:
\begin{eqnarray*}
	&&{\rm do}\ k=1,N\\
		&&\qquad (A_k^{(t+1)},H_k^{(t+1)}) =f\{(A_1^{(t+1)},H_1^{(t+1)}),... \\
		&&\qquad\qquad ..,(A_{k-1}^{(t+1)},H_{k-1}^{(t+1)}),(A_k^{(t)},H_k^{(t)}),...,(A_n^{(t)},H_n^{(t)})\}
	\\
	&&{\rm end\ do}
\label{simultaneous_update}
\end{eqnarray*}
where the function $ f $ represents the operation in the right side of Eq. (\ref{cavity_dist_update}).
To stably obtain right results, we need to employ this sequential update strategy or other update strategy which possesses similar characteristics with the sequential update strategy.
For example, if the parallel update strategy, i.e.,
\begin{eqnarray*}
	&&{\rm do}\ k=1,N\\
		&&\qquad (A_k^{(t+1)},H_k^{(t+1)}) =f\{(A_1^{(t)},H_1^{(t)}),...,(A_n^{(t)},H_n^{(t)})\}
	\\
	&&{\rm end\ do}
\label{sequential_update}
\end{eqnarray*}
is used, then the simulation may not converge.

\section{Calculation methodology for eigenvalue/vector}
\label{Calculation methodology}

\subsection{Power method}\label{ss_power_method}

We use the power method to calculate the first eigenvalue and eigenvector of a given adjacency matrix.  This method is well known to evaluate the first eigenvalue/vector numerically in practice.  Assume that $ \bmA $ is the $ n\times n$ matrix having $ n $ distinct eigenvalues $ \lambda_1,\ \lambda_2,...,\lambda_n $. The eigenvalues are ordered in decreasing magnitude, i.e., $ |\lambda_1|> | \lambda_2|\ge\cdots \ge|\lambda_n| $.  We calculate the following sequential equations from an appropriately chosen initial vector $ \bmx_0 $.
\begin{eqnarray*}
\label{power_method}
	\bmy_{k} &=&  \bmA\, \bmx_k,
	\\
	\bmx_{k+1} &=& \frac{1}{c_k}\, \bmy_k,
\end{eqnarray*}
where $ c_k $ is an appropriate number to avoid the divergence of $ |\bmx_k| $. If we substitute the value of $ |\bmy_k| $ into $ c_k $, then $ c_k $ and $ \bmx_k $ converge to the first eigenvalue $ \lambda_1 $ and the first eigenvector $ \bmv_1 $, respectively, i.e.,
\begin{eqnarray*}
\label{power_method_conv}
	\lim_{k\to\infty} \bmx_k &=& \bmv_1,
	\\
	\lim_{k\to\infty} c_k &=& \lambda_1.
\end{eqnarray*}

In our case, the adjacency matrix $ \bmJ $ has eigenvalues $ \lambda_1',\ \lambda_2',...,\lambda_n' $ which are ordered as $ \lambda_1'> \lambda_2'\ge \cdots \ge \lambda_n' $.  Then, $ \bmJ $ might have a negative eigenvalue whose absolute value is larger than the largest eigenvalue, i.e. $ \lambda_1' < |\lambda_k'| $.  For this reason, we reconstruct the following matrix to evaluate the eigenvalue/vector of $ \bmJ $.
\begin{eqnarray*}
\label{powerA=J+I}
	\bmA = \bmJ + \eta\, \bmI,
\end{eqnarray*}
where $ \bmI $ is the identity matrix and $ \eta $ is a certain positive number that satisfies $ \eta > \min{0,-\lambda_1'-\lambda_k'}/2 $.  Additionally, the eigenvalues of $ \bmA $ become  $ \lambda_1' + \eta,\ \lambda_2 + \eta ',...,\lambda_n + \eta ' $.

\subsection{Scaling of $ \varLambda $}
To find the value $ \varLambda $ for $ N=\infty $ using the power method, we assume that the eigenvalue $ \varLambda $ and the size of the matrix $N $ are related as follows:
\begin{eqnarray} 
	\varLambda(N) = \exp (A\, N^{-\beta}+B)
\label{assumptionLambda(N)=..}
\end{eqnarray}
where for each $ \varDelta$, $ A $ is a constant and $ B$ and $ \beta $ are positive constants. We can evaluate $\varLambda(\infty) = \exp (B)$.
We found the above equation heuristically from several possible candidates and there is not theoretical validity for the equation so far.  To evaluate the values $ A,\ B$ and $\beta $, we take the logarithm of the above equation:
\begin{eqnarray*} 
	\log \varLambda(N) = A\, N^{-\beta}+B
\end{eqnarray*}
and using the linear regression, we find the optimal values of $ A,\, B$ and $ \beta $.
This relationship holds very well in some cases, however, the relation does not hold for all the cases.

\subsection{Normalization of eigenvalues}
In this section, we describe the normalization method of the first eigenvector $ \bmv $ and the values of $ v_i\ (= H_i/A_i) $ in the cavity method.  It is necessary to normalize these for better comparison.  There are many normalization methods and one of the most common methods is to set the value of the variance equals 1, i.e.,
\begin{eqnarray}
\label{def.scale-matrix-cavity} 
	\frac{1}{N}\sum_i v_i{}^2 = 1,
\end{eqnarray}
where $ N $ is the size of the vector or the number of population in the population dynamical method.  We use the above normalization in Fig.~\ref{fig4-8-09eigenvector-scale} (b), Fig.~\ref{fig4-8-09eigenvector} of $ \varDelta = 0.8,\ 1.0$, Fig.~\ref{fig1d4eigenvector}, Fig.~\ref{fig1d8eigenvector} and other similar figures below.

The above normalization is not available when the tail of the density decreases as
\begin{eqnarray}
	d(x) \sim x^{-(1+\alpha)}, \qquad (1<\alpha\le2),
\end{eqnarray}
because the density function does not have a finite variance.  In this case, we need to use another normalization method such as
\begin{eqnarray}
\label{def.normalize-matrix-cavity_abs} 
	\frac{1}{N}\sum_i |v_i| = 1.
\end{eqnarray}
We use this normalization in Figs.~\ref{fig4-8-09eigenvector-scale} (a) and (c), Fig.~\ref{fig4-8-09eigenvector} of $ \varDelta = 0.0,\ 0.3,\ 0.6$, and other similar figures below.

\section{Result}\label{Results}
In this section, we show the results for the models whose network nodes have two different types of degrees and whose degrees distribution is the Poissonian distribution.  In addition, we show the results for the Laplacian matrices in the Appendix.

\subsection{2-DTD network}

\begin{figure}
\rotatebox{90}{\hspace{35mm}\rotatebox{0}{$ \varLambda $}}\hspace{-4mm}
\includegraphics[width=7cm]{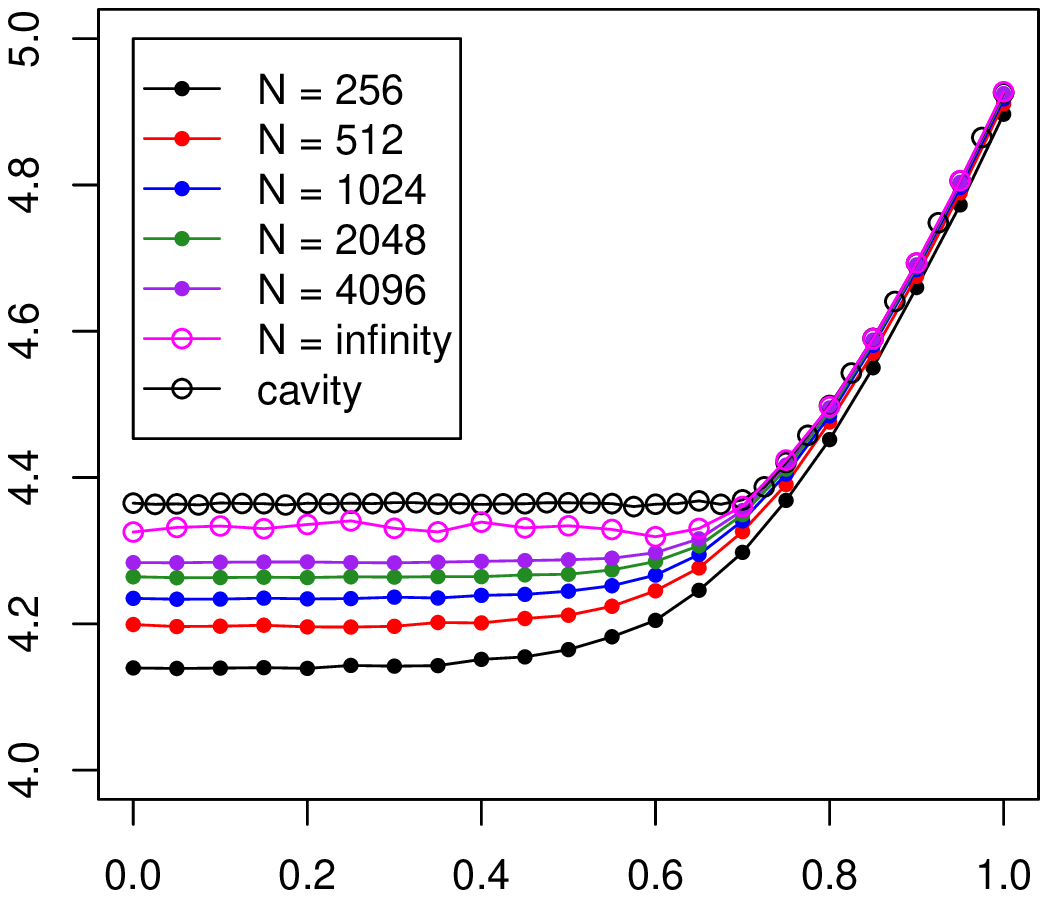}\vspace{-10mm}\\
\hspace{10mm} $ \varDelta$
\caption{First eigenvalue $ \varLambda $ versus $ \varDelta $ for 2-DTD model.  Networks are constructed by 4 and 8 degrees of links in the ratio of 0.9:0.1.}
\label{figure2d4-8-09Lambda}
\end{figure}

We show the results for networks that are constructed by the 4 or 8 degrees of connections with a ratio of 0.9:0.1.  Figure \ref{figure2d4-8-09Lambda} shows the results of the first eigenvalues evaluated by the cavity and power methods.  For each $ \varDelta$, the values of $ \varLambda $ are increasing with the system size $ N $.  To evaluate the $ \varLambda $, we take the average of 2000 configurations, i.e., we calculate 2000 first eigenvalues from 2000 different adjacency matrices.

\begin{figure}
\begin{minipage}{80mm}
(a) $ \varDelta =0.0 $\vspace{-10mm}\\
\rotatebox{90}{\hspace{25mm}\rotatebox{0}{$ \log \varLambda $}}\hspace{-5mm}
\includegraphics[width=6cm]{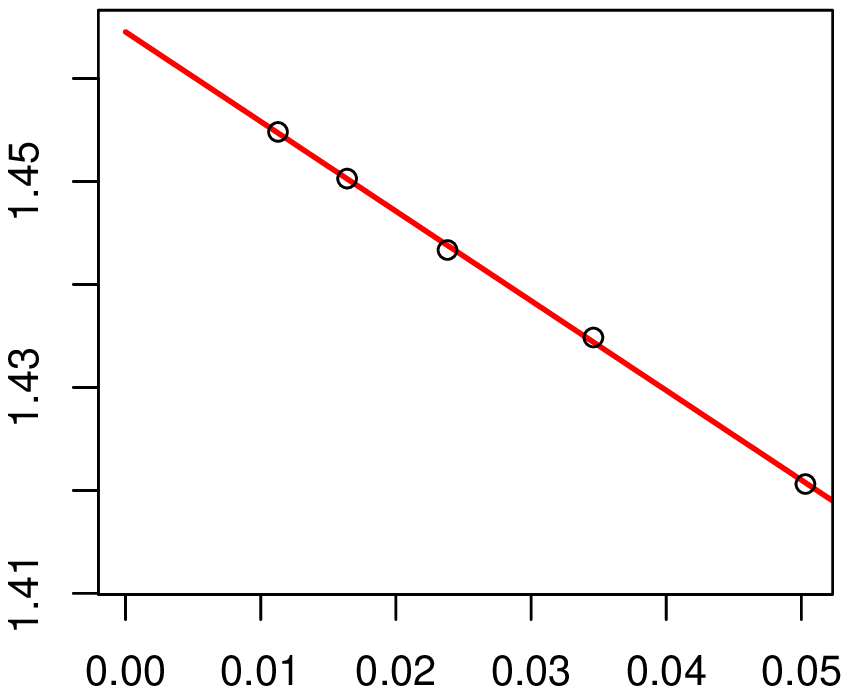}\vspace{-10mm}\\
\hspace{10mm} $N^{-\beta}$
\end{minipage}
\begin{minipage}{80mm}
(b) $ \varDelta =0.8 $\vspace{-10mm}\\
\rotatebox{90}{\hspace{25mm}\rotatebox{0}{$ \log \varLambda $}}\hspace{-5mm}
\includegraphics[width=6cm]{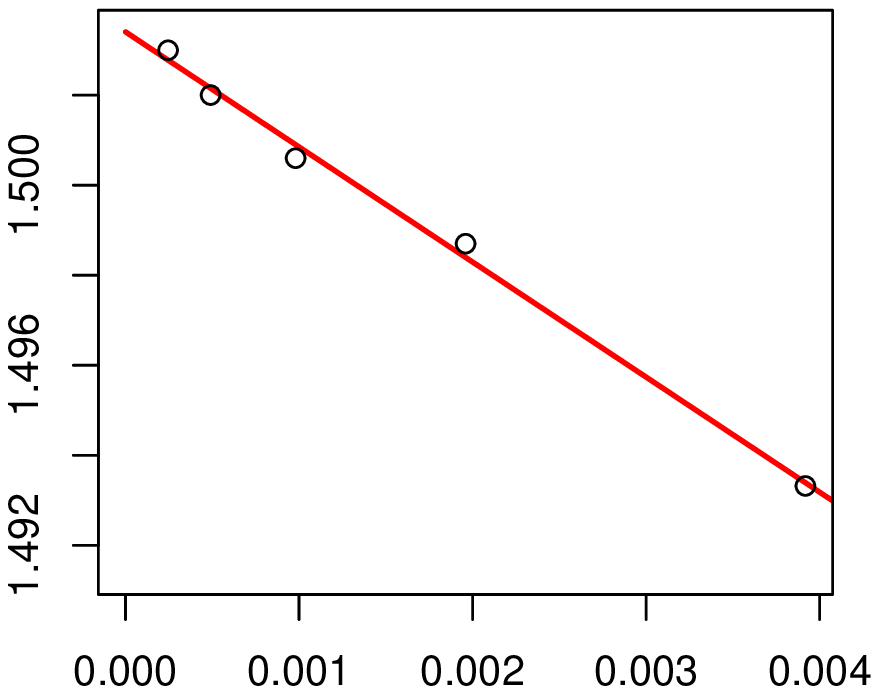}\vspace{-10mm}\\
\hspace{10mm} $N^{-\beta}$
\end{minipage}
\caption{Results of the linear regression to find $\varLambda(\infty)$. The lines represent $\log \varLambda(N) = A\, N^{-\beta}+B$. In (a), $ A=-0.870 $, $B=1.465$, $\beta=0.539$.  In (b), $ A=-2.556 $, $B=1.503$, $\beta=0.999$.}
\label{fig4-8-09beta}
\end{figure}

\begin{figure}
\rotatebox{90}{\hspace{35mm}\rotatebox{0}{$ \beta $}}\hspace{-5mm}
\includegraphics[width=7cm]{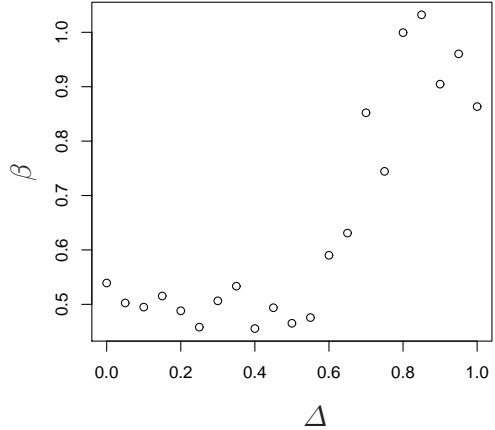}\vspace{-10mm}\\
\hspace{10mm} $ \varDelta $
\caption{Values of $ \beta $.}
\label{fig4-8-09beta-delta}
\end{figure}

To estimate the first eigenvalue $ \varLambda $ for $ N=\infty $ on the basis of the results of the power method, we assume that the eigenvalue $ \varLambda $ and the size of the matrix $N $ satisfy the scaling relation of Eq.~(\ref{assumptionLambda(N)=..}).  The results fit very well as shown in Fig.~\ref{fig4-8-09beta}.  Figure \ref{fig4-8-09beta-delta} shows the values $\beta$ evaluated by the above method.  We estimate the critical value $ \varDelta_{\rm c}=0.611$, using the scaling method in Ref.~\onlinecite{Kabashima}.  Corresponding with this critical value, the values of $\beta$ are around 0.5 in the region of $ \varDelta =0.0-0.6 $ and around 0.9 in the region of $ \varDelta =0.7-1.0 $. 

The results of the cavity method in Fig.~\ref{figure2d4-8-09Lambda} resemble those obtained by the power method, although the results of $ \varLambda $ using the cavity method are greater than those obtained by the power method with the scaling correction.  The reason for this discrepancy is not clear but it might be that the results from the scaling correction are underestimated or the results of the cavity method are overestimated for some reason.  As is well known for this type of problem, the finite size correction may be necessary for the finite size scaling.  The line of $ \varLambda $ calculated by the cavity method becomes almost flat at a little less than around $ \varDelta =0.7 $.  This value is not corresponding with the critical value $ \varDelta_{\rm c}=0.611$ which we described in Ref.~[\onlinecite{Kabashima}].

\begin{figure}
\begin{minipage}{80mm}
(a)$\varDelta = 0.0$\vspace{-10mm}\\
\rotatebox{90}{\hspace{30mm}\rotatebox{0}{$ \log d(v_i) $}}\hspace{-5mm}
\includegraphics[width=7cm]{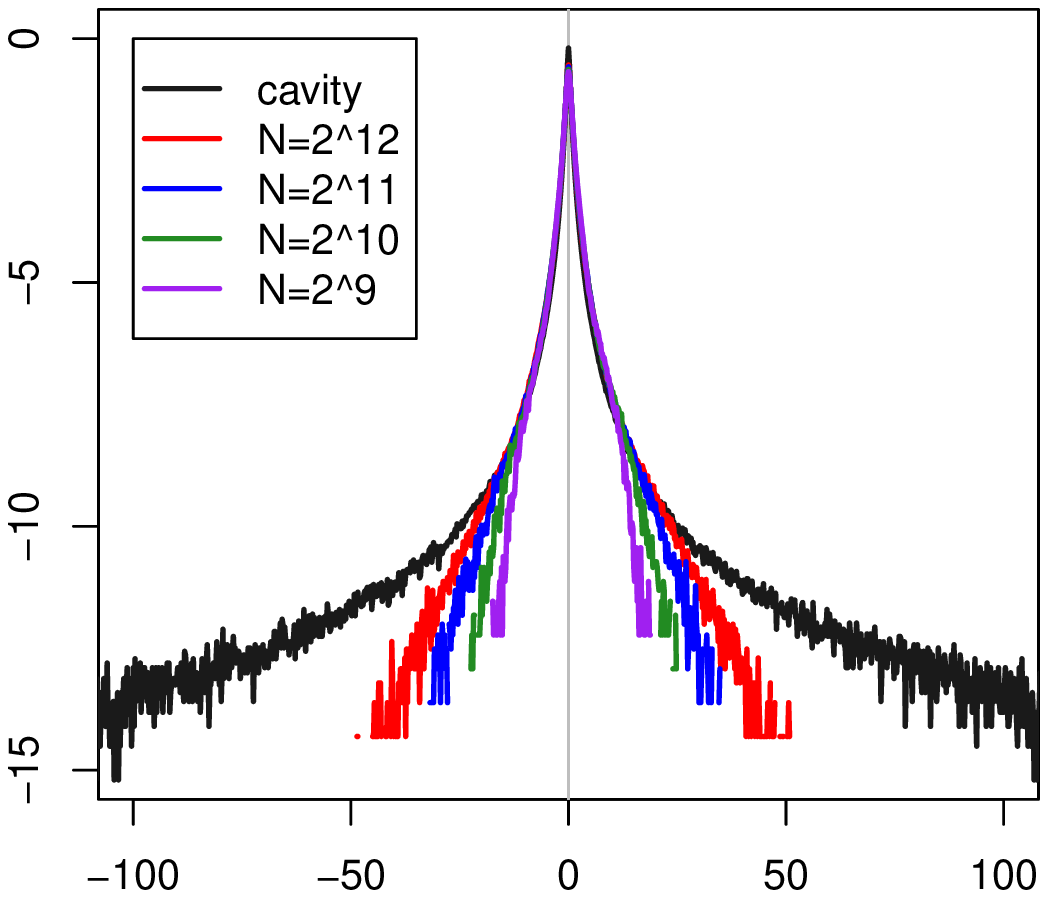}\vspace{-12mm}\\
\hspace{10mm} $ v_i $ \vspace{-62mm}\\
\hspace{35mm}\includegraphics[width=3cm]{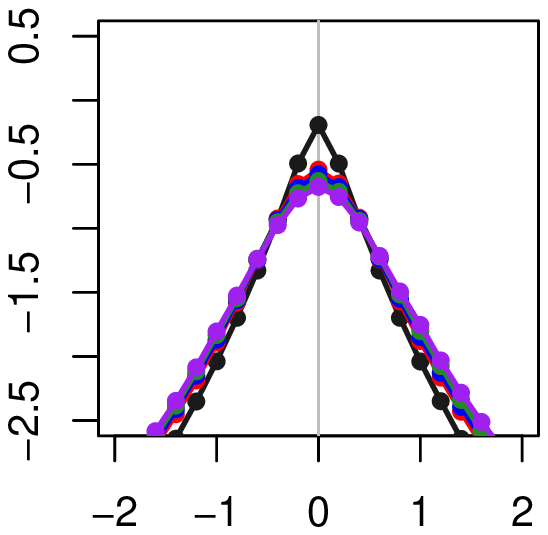}\vspace{-12mm}\vspace{39mm}\\
\end{minipage}
\begin{minipage}{80mm}
(b)$\varDelta = 0.8$\vspace{-10mm}\\
\rotatebox{90}{\hspace{30mm}\rotatebox{0}{$ \log d(v_i) $}}\hspace{-5mm}
\includegraphics[width=7cm]{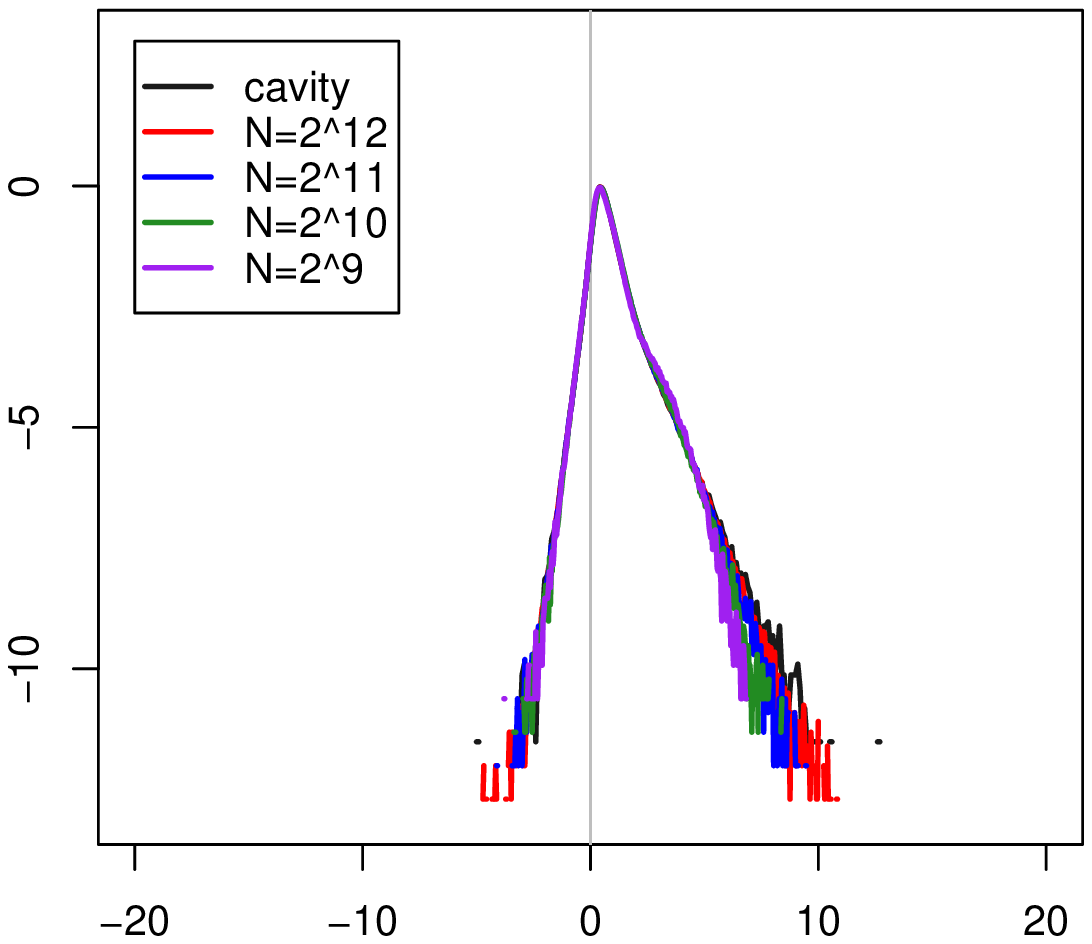}\vspace{-12mm}\\
\hspace{10mm} $ v_i $ \vspace{-62mm}\\
\hspace{37mm}\includegraphics[width=2.8cm]{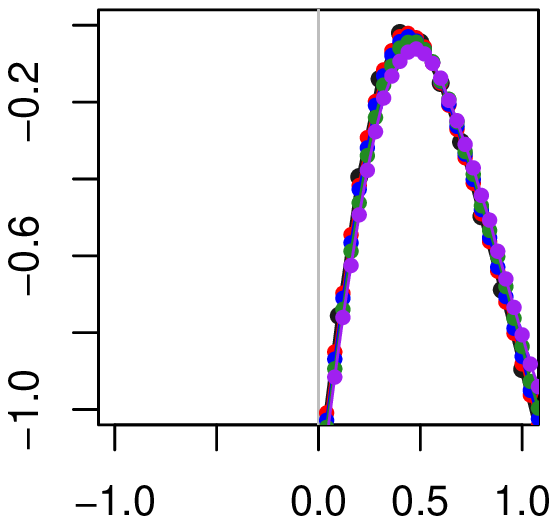}\vspace{-12mm}\vspace{39mm}\\
\end{minipage}
\vspace{5mm}

\begin{minipage}{80mm}
(c)$\varDelta = 0.0$\vspace{-10mm}\\
\rotatebox{90}{\hspace{25mm}\rotatebox{0}{$ \log  \int_{v_i}^\infty d(x) dx $}}\hspace{-5mm}
\includegraphics[width=7cm]{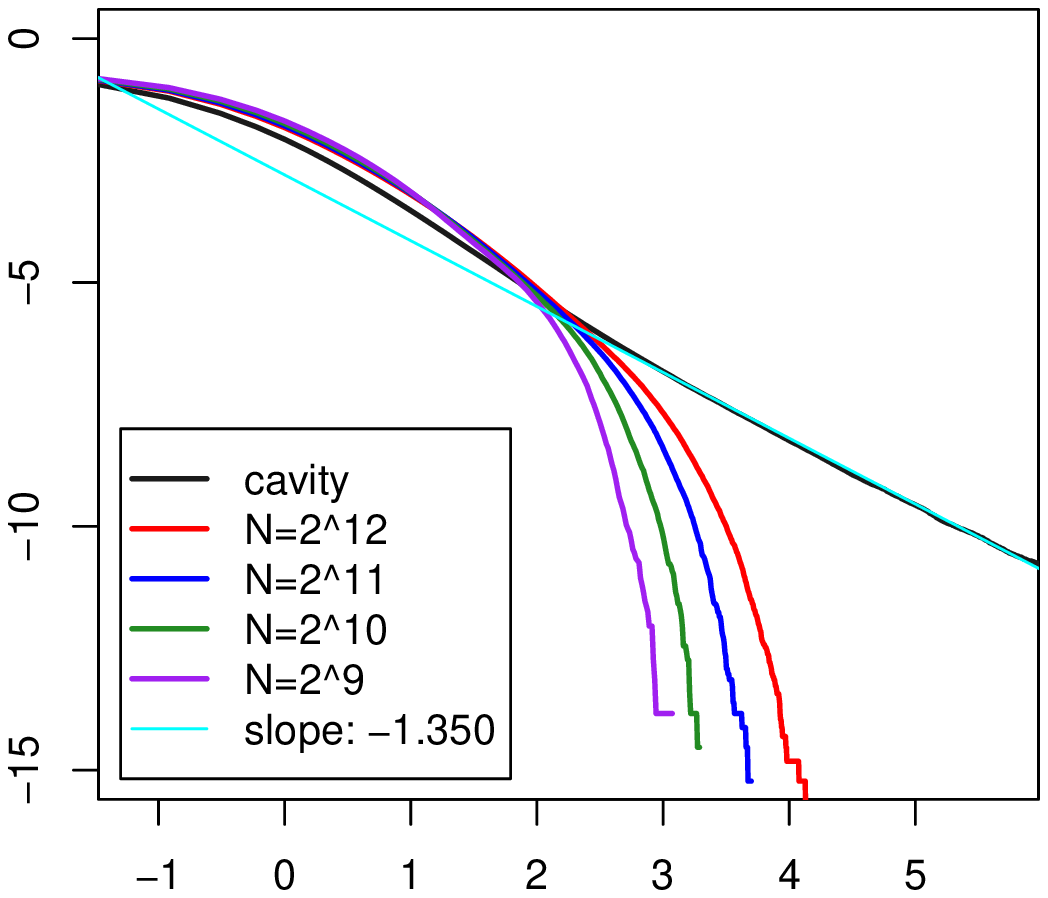}\vspace{-12mm}\\
\hspace{10mm} $ \log v_i $\\
\end{minipage}
\begin{minipage}{80mm}
(d) Convergence\vspace{-10mm}\\
\rotatebox{90}{\hspace{30mm}\rotatebox{0}{$ \log S $}}\hspace{-5mm}
\includegraphics[width=7cm]{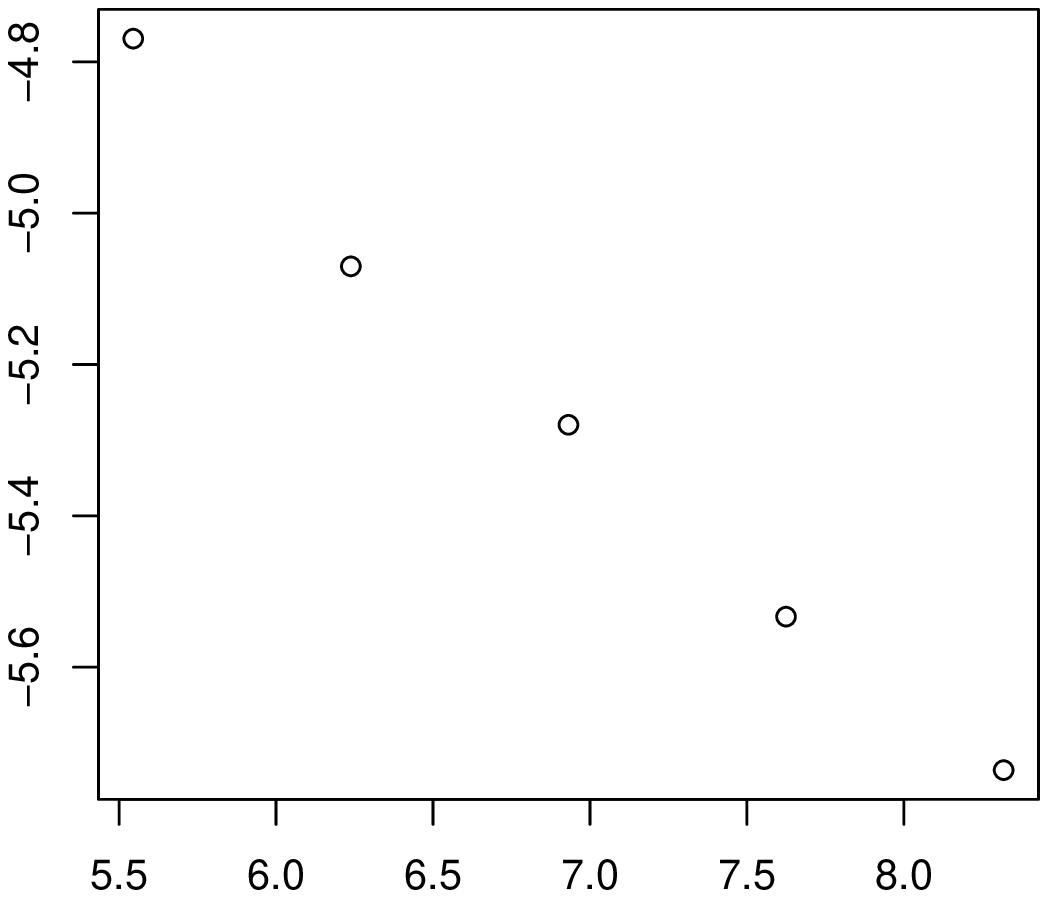}\vspace{-12mm}\\
\hspace{5mm} $ \log N $\\
\end{minipage}
\caption{Scaled density of the first eigenvector $ \bmv $ for 2-DTD model.  Networks are constructed by 4 or 8 degrees of links with the ratio of 0.9:0.1.  (a) is for $\varDelta = 0.0$ and is normalized as $ \sum_i |v_i| = N $.  An inset is its magnification around $ v_i=0 $ region.  (b) is for $\varDelta = 0.8$ and normalized as $ \sum_i v_i^2 = N $.  (c) is the log-log plot of the cumulative distribution of (a) for the region of $v_i >0 $.  (d) shows the convergence to the result of the cavity method assessed by $S$ in the limit of $ N \to \infty $.}
\label{fig4-8-09eigenvector-scale}
\end{figure}

Now, we confirm that the density functions evaluated by the first eigenvector of the adjacency matrix and the cavity method are consistent unless the finite size effect.  We use a notation $ d(v_i) $ to represent both the density function of the first eigenvectors $ \bmv $ and the values $ v_i\ (= H_i/A_i) $ in the cavity method.  Figures \ref{fig4-8-09eigenvector-scale} (a) and (b) show the density functions $ d(v_i) $ of the model whose network nodes have two different types of degrees, that is the degree 4 and 8 with a ratio of 0.9:0.1.  Figure \ref{fig4-8-09eigenvector-scale}(a) is $ \varDelta =0.0 $ and (b) is $ \varDelta =0.8 $.  We compare the density functions for four different system sizes $ N=2^9,\, 2^{10},\, 2^{11},\, 2^{12} $ and the results of the cavity method in each figure.  There are some differences between smaller and larger $ \varDelta $ whose boundary is around the critical point $ \varDelta_c $.  For smaller $ \varDelta $, for example $ \varDelta=0.0 $, there are discrepancies around $ v_i = 0 $ and at the tail of the density function (see (c) and the inset of (a)).  However, for larger $ \varDelta $, the shapes of the density functions are similar for all $N$ and similar to the result of the cavity method, see Fig~\ref{fig4-8-09eigenvector-scale}(b) $ \varDelta=0.8 $ for example.

In Fig.~\ref{fig4-8-09eigenvector-scale}(a), $ \varDelta = 0.0 $, the tail of the densities based on the power method is shorter than that based on the cavity method.  This is reasonable because the eigenvector of the finite size matrix with the finite entries must have a cut-off.  The discrepancy around the $ v_i = 0 $ region is also caused because the size of the matrix is finite (see the inset of Fig.~\ref{fig4-8-09eigenvector-scale}(a).)  Figure~\ref{fig4-8-09eigenvector-scale}(c) shows the log-log plot of the cumulative distribution of Fig.~\ref{fig4-8-09eigenvector-scale} (a) for the region of $v_i>0 $.  The slope of a straight line in the figure is $-1.350$ and this represents $ \alpha = 1.350 $.

To confirm the convergence of the density functions, we define the following value:
\begin{eqnarray}
	\label{convergence-assess}
	S = \sum_i (d^{(A)}_i-d^{\rm (cav)}_i)^2,
\end{eqnarray}
where $ d^{(A)}_i $ and $ d^{\rm (cav)}_i $ represent an appropriately partitioned density of $ v_i $.  In this study, the partition of the interval of $ v_i $ was taken as 0.2 to calculate $S$.  Figure \ref{fig4-8-09eigenvector-scale}(d) shows the values of $S$ for each $N$.    We find that the density $ d^{(A)}_i $ converge monotonically to $ d^{\rm (cav)}_i $ with increasing the network size.

\begin{figure}
\begin{minipage}{80mm}
(a) $ N=2^{12} $ \vspace{-10mm}\\
\rotatebox{90}{\hspace{35mm}\rotatebox{0}{$ \log d(v_i) $}}\hspace{-4mm}
\includegraphics[width=7cm]{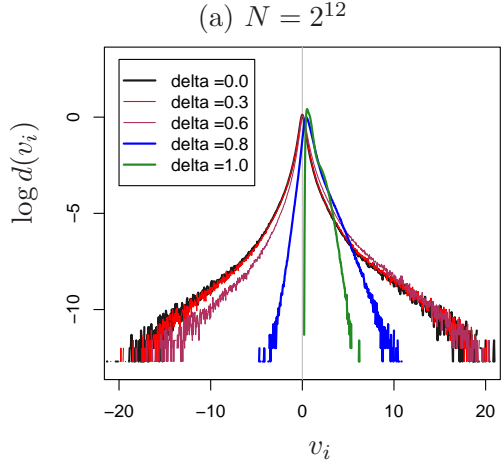}\vspace{-12mm}\\
\hspace{10mm} $ v_i $\\
\end{minipage}
\caption{Densities of the first eigenvector $ \bmv $ for 2-DTD model.  Networks are constructed by the degree 4 and 8 of links with a ratio of 0.9:0.1.  The sizes of the matrices are $ N=2^{12} $.  The densities are evaluated from two thousand samples.}
\label{fig4-8-09eigenvector}
\end{figure}

Figure \ref{fig4-8-09eigenvector} shows the density functions of the first eigenvectors for the network whose nodes have two different types with respect to the degrees. Using $ N=2^{12} $, we compare the density functions of five different $ \varDelta$, i.e., $ \varDelta = 0.0,\ 0.3,\ 0.6,\ 0.8,\ 1.0$.  We find that the density function has a fat tail in the region $ \varDelta < \varDelta_c $, although not in the region $ \varDelta < \varDelta_c $.  The results obtained from the cavity method are similar to these results, unless there exist some discrepancies, which we already mentioned. 

\begin{figure}
\begin{minipage}{80mm}
$ N=2^{12} $ \vspace{-10mm}\\
\rotatebox{90}{\hspace{35mm}\rotatebox{0}{$ \log d(v_i) $}}\hspace{-4mm}
\includegraphics[width=7cm]{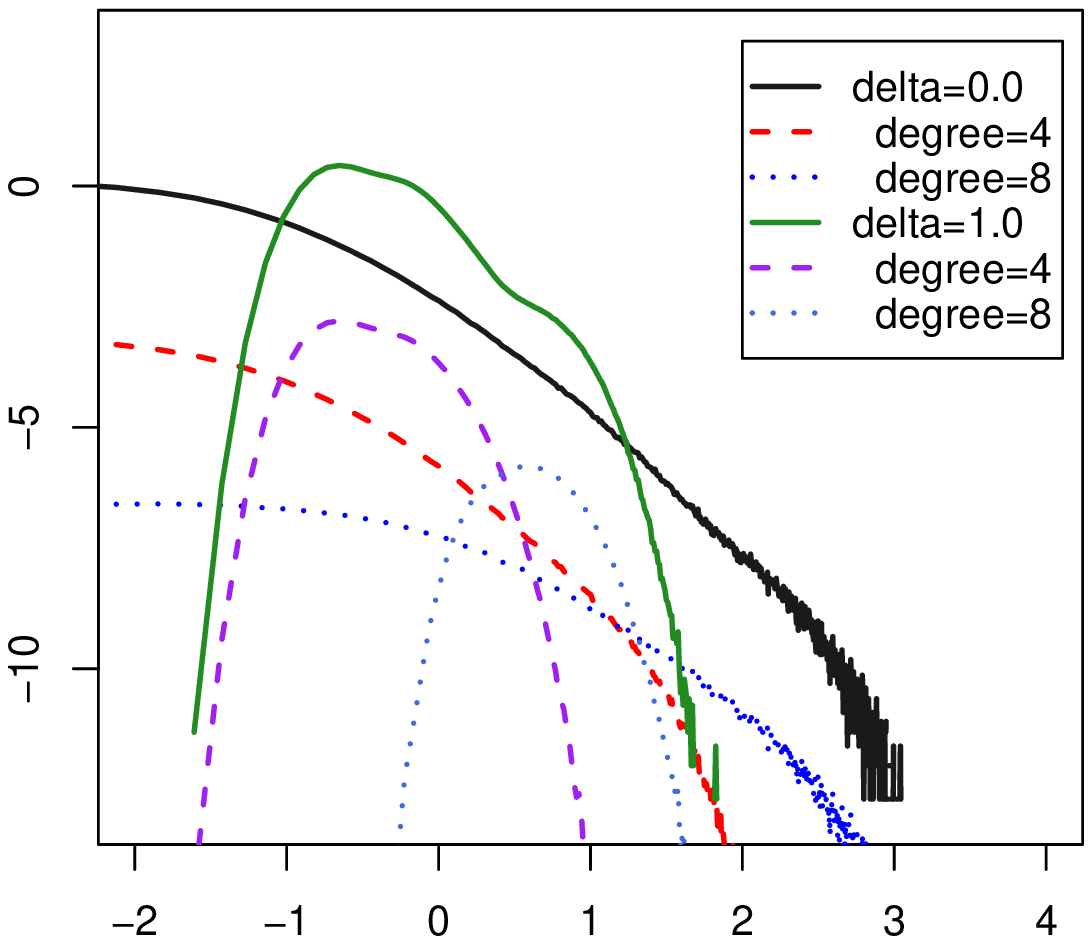}\vspace{-12mm}\\
\hspace{10mm} $ \log v_i $\\
\end{minipage}
\caption{Contribution to density from two different degrees.  Networks are constructed by degree 4 and 8 of links with a ratio of 0.9:0.1. }
\label{fig2d4-8-09eigenvector-d}
\end{figure}

In Fig.~\ref{fig2d4-8-09eigenvector-d}, we show that the contribution to the density from the two different degrees for $ \varDelta\ =0.0$, and $ 1.0 $.  We find that the larger values of $ v_i $ are constructed mainly of the nodes whose degree is the larger and the smaller values of $ v_i $ are constructed mainly of the nodes whose degree is the smaller, and vice versa.

\begin{figure}
\begin{minipage}{80mm}
\rotatebox{90}{\hspace{30mm}\rotatebox{0}{$ \log d(v_i) $}}\hspace{-4mm}
\includegraphics[width=7cm]{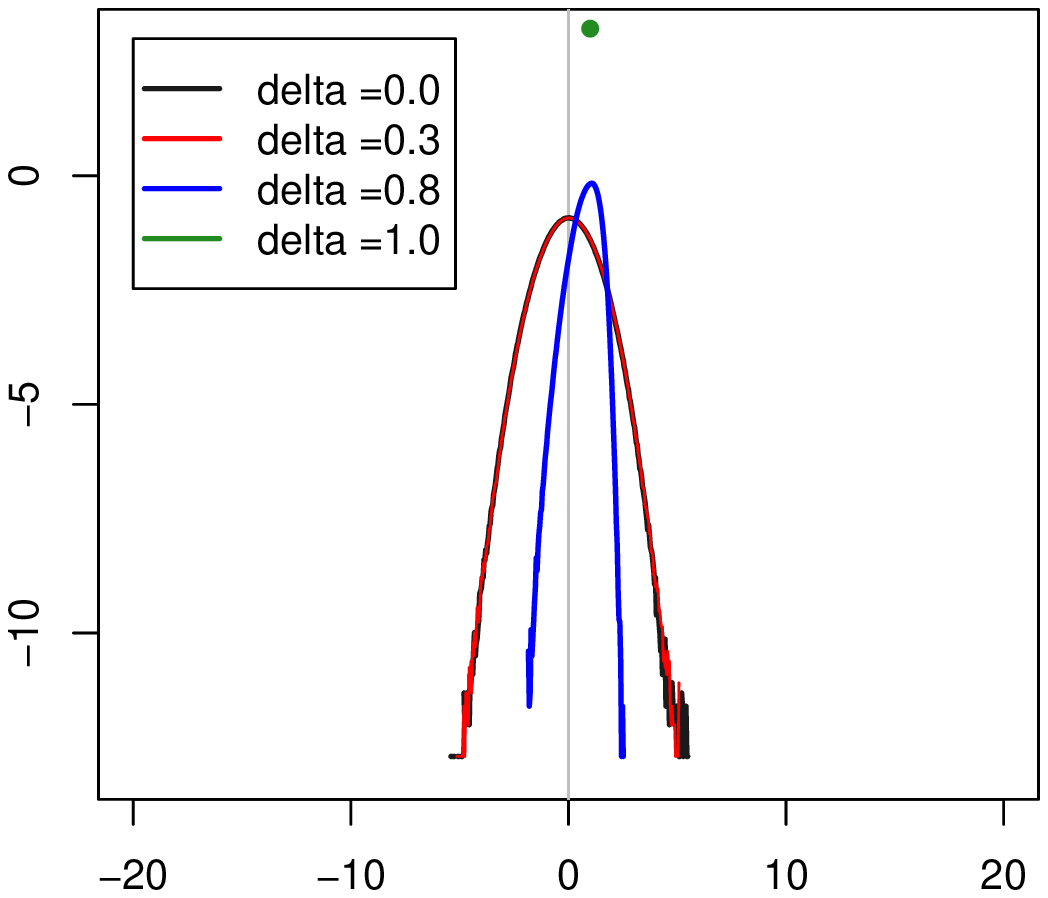}\vspace{-13mm}\\
\hspace{10mm} $ v_i $\vspace{-61mm}\\
\hspace{38mm}\includegraphics[width=2.8cm]{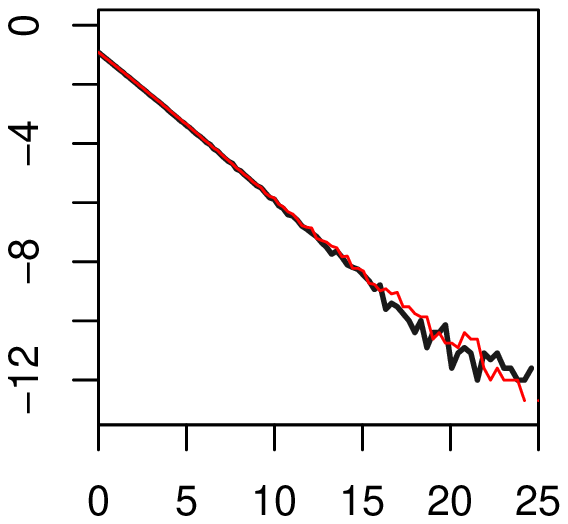}\vspace{-9mm}\\
\hspace{45mm} $ v_i{}^2 $ \vspace{30mm}\\
\end{minipage}
\caption{Densities of the first eigenvector $ \bmv $.  Networks are 4-regular graph, only.  The sizes of matrices are $ N=2^{12} $.  $x$-axes is $ v_i^2 $ in the inset.}
\label{fig1d4eigenvector}
\end{figure}

\begin{figure}
\begin{minipage}{80mm}
\rotatebox{90}{\hspace{35mm}\rotatebox{0}{$ \log d(v_i) $}}\hspace{-4mm}
\includegraphics[width=7cm]{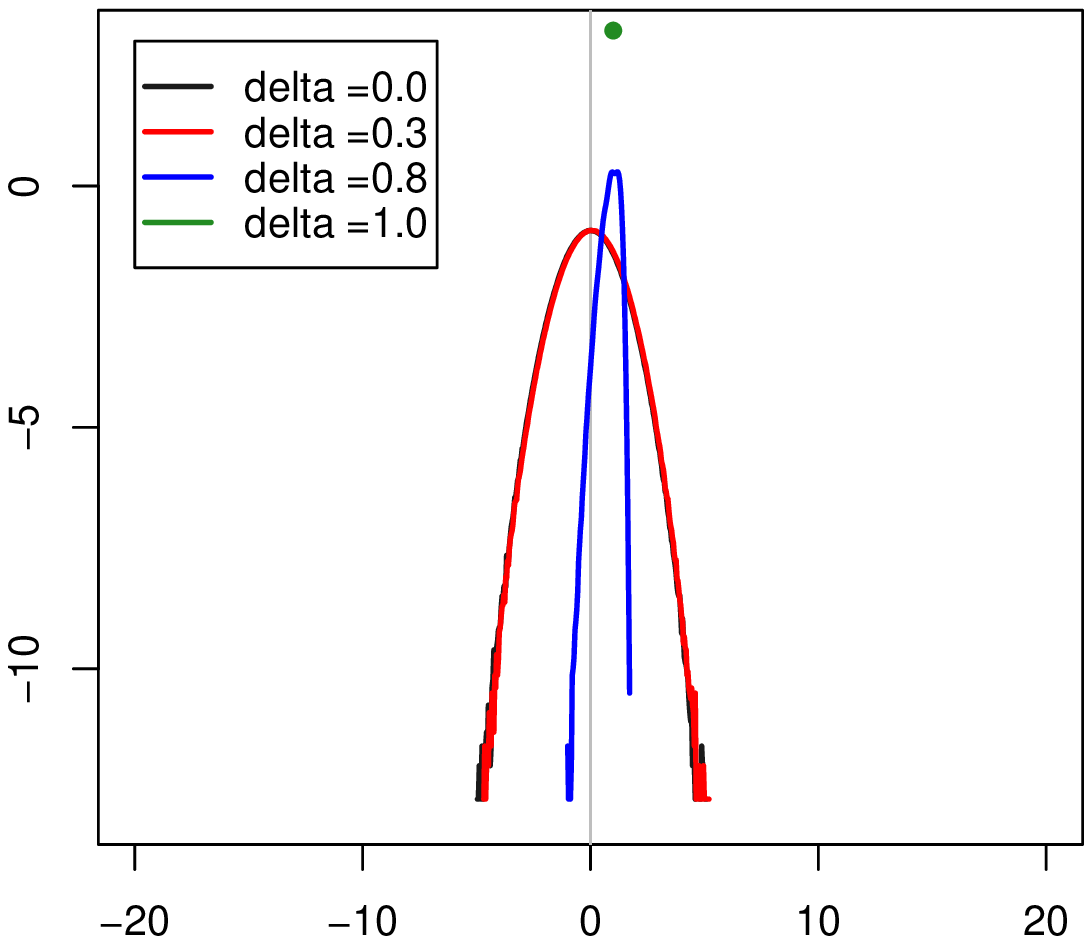}\vspace{-12mm}\\
\hspace{10mm} $ v_i $ 
\vspace{-61mm}\\
\hspace{38mm}\includegraphics[width=2.8cm]{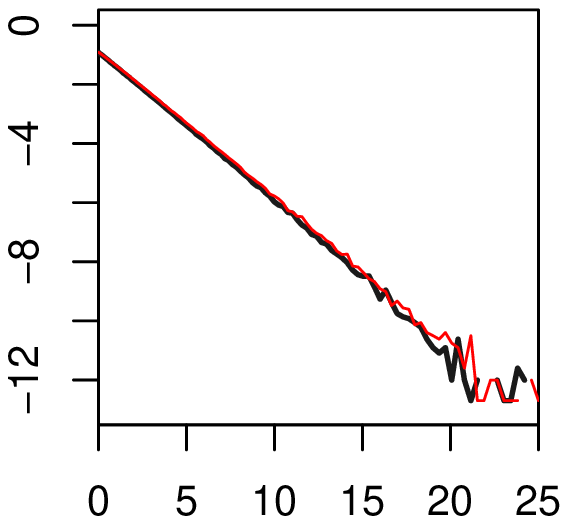}\vspace{-10mm}\\
\hspace{40mm} $ v_i^2 $ \vspace{30mm}\\
\end{minipage}
\caption{Densities of the first eigenvector $ \bmv $.  Networks are 8-regular graph.  The sizes of the matrices $ N=2^{12} $.  $x$-axis is $ v_i^2 $ in the inset.}
\label{fig1d8eigenvector}
\end{figure}

For comparison with the above results, in Fig.~\ref{fig1d4eigenvector}, we show results of another network where all the nodes are degree 4.  The shapes of the densities are different from those of 2-DTD model (e.g. Fig.~\ref{fig4-8-09eigenvector}.)  The tail of the density function $ d(v_i) $ is similar to the Gaussian distribution, whereas the power law for the 2-DTD model for small $ \varDelta $.  In the case that the degree of all the network nodes is only one type and degree 8, the results on the density function $ d(v_i) $ is similar to those of the case in which all the nodes are degree 4, see Fig.~\ref{fig1d8eigenvector}.  From these results, we conclude that the heavy tail of the density function, which we saw in Fig.~\ref{fig4-8-09eigenvector}, is produced when the network has two different types of nodes with respect to the degree.

\begin{figure}
\begin{minipage}{80mm}
(a)\vspace{-10mm}\\
\rotatebox{90}{\hspace{35mm}\rotatebox{0}{$ \log d(v_i) $}}\hspace{-4mm}
\includegraphics[width=7cm]{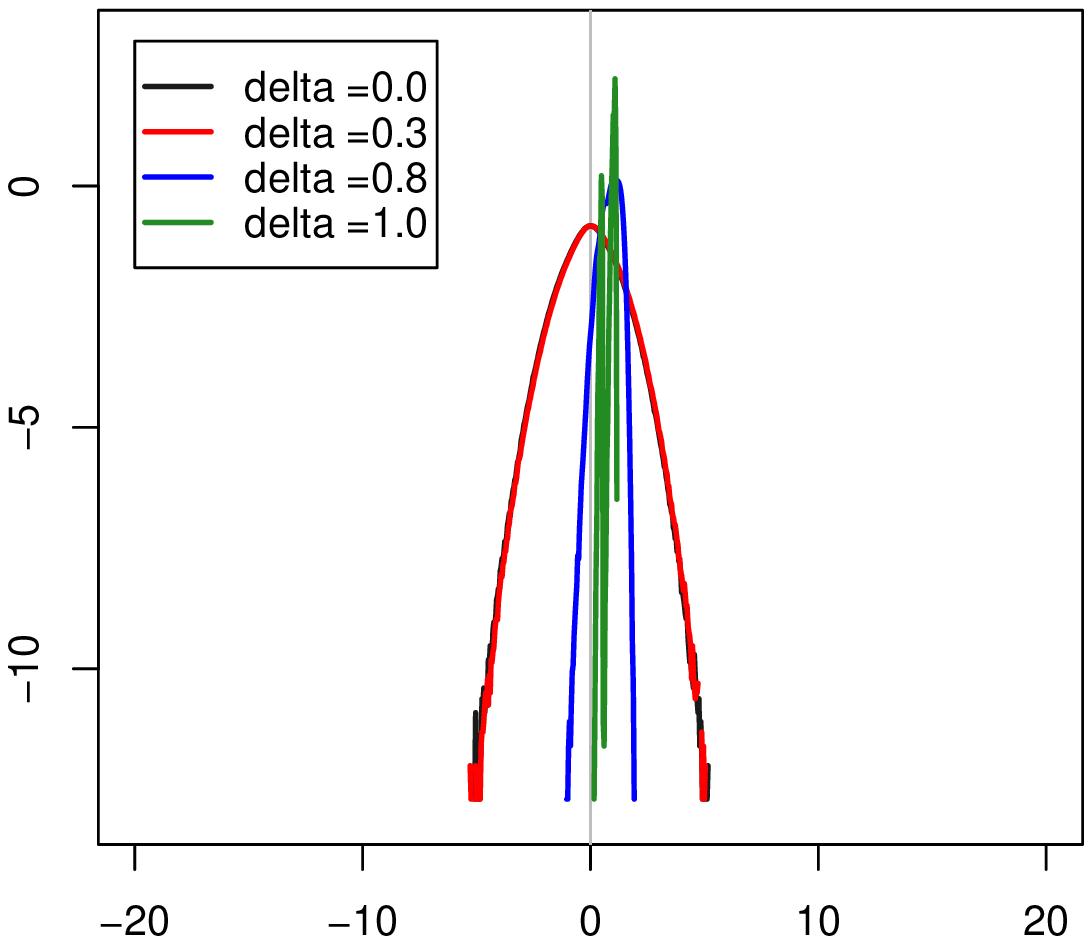}\vspace{-12mm}\\
\hspace{10mm} $ v_i $\vspace{-62mm}\\
\hspace{38mm}\includegraphics[width=2.8cm]{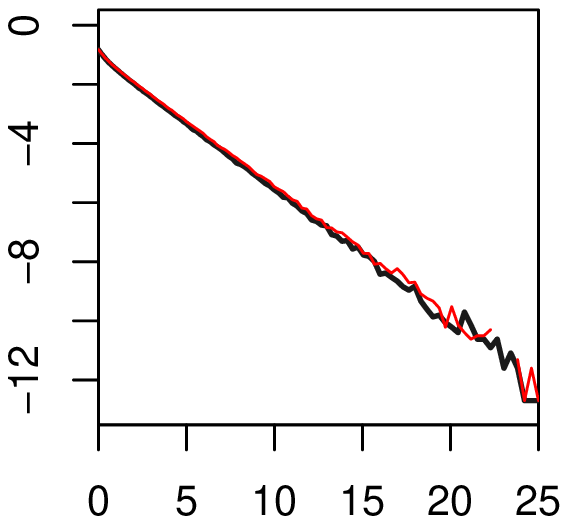}\vspace{-10mm}\\
\hspace{41mm} $ v_i^2 $ \vspace{30mm}\\
\end{minipage}
\begin{minipage}{80mm}
(b)\vspace{-10mm}\\
\rotatebox{90}{\hspace{35mm}\rotatebox{0}{$ \log d(v_i) $}}\hspace{-4mm}
\includegraphics[width=7cm]{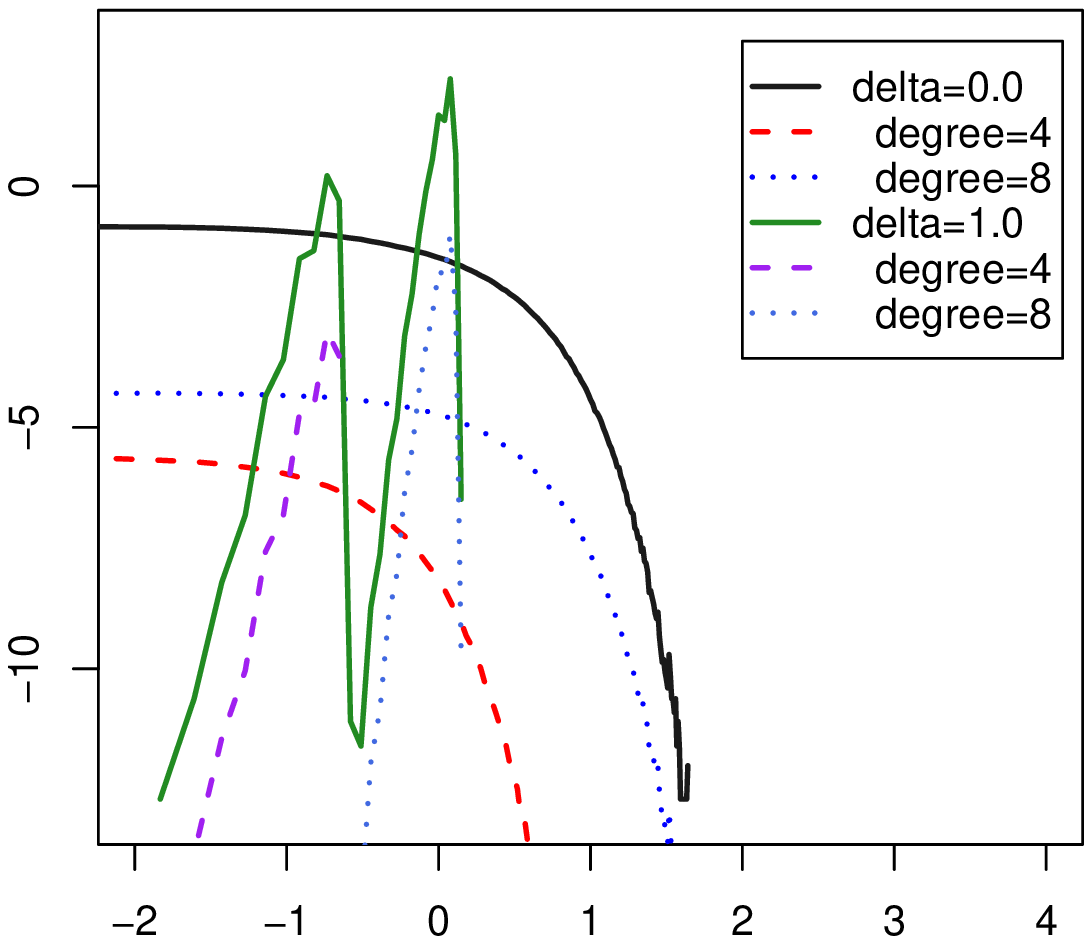}\vspace{-12mm}\\
\hspace{10mm} $ \log v_i $\\
\end{minipage}
\caption{Densities of the first eigenvector $ \bmv $ for the model whose network nodes are constructed by the 4 or 8 degrees of links with a ratio of 0.1:0.9.  The sizes of the matrices are $ N=2^{12} $. (a) shows results of four different $\varDelta$.  (b) shows the contribution to the density from two different degrees. }
\label{fig2d48-01eigenvector}
\end{figure}

Figure \ref{fig2d48-01eigenvector} shows the results of the model whose network nodes have two different types of degrees.  The networks are connected by the nodes of degree 4 and 8 and the ratio is 0.1:0.9.  Compared to the case in which the ratio is 0.9:0.1, see Fig.~\ref{fig4-8-09eigenvector}, the tail of the density function in Fig.~\ref{fig2d48-01eigenvector} is similar to the Gaussian distribution which we observe when the network nodes are only one type of degree.

Another feature is that there are two peaks for the case of $ \varLambda =1.0 $, because the network has two different type of degrees.  The peak at the smaller $v_i $ mainly comes out from the nodes with degree 4 and the peak at the larger $v_i $ mainly originates from the nodes with degree 8, and vice versa, see Fig.~\ref{fig2d48-01eigenvector} (b).

\begin{figure}
\begin{minipage}{80mm}
(a)\vspace{-10mm}\\
\rotatebox{90}{\hspace{35mm}\rotatebox{0}{$ \log d(v_i) $}}\hspace{-4mm}
\includegraphics[width=7cm]{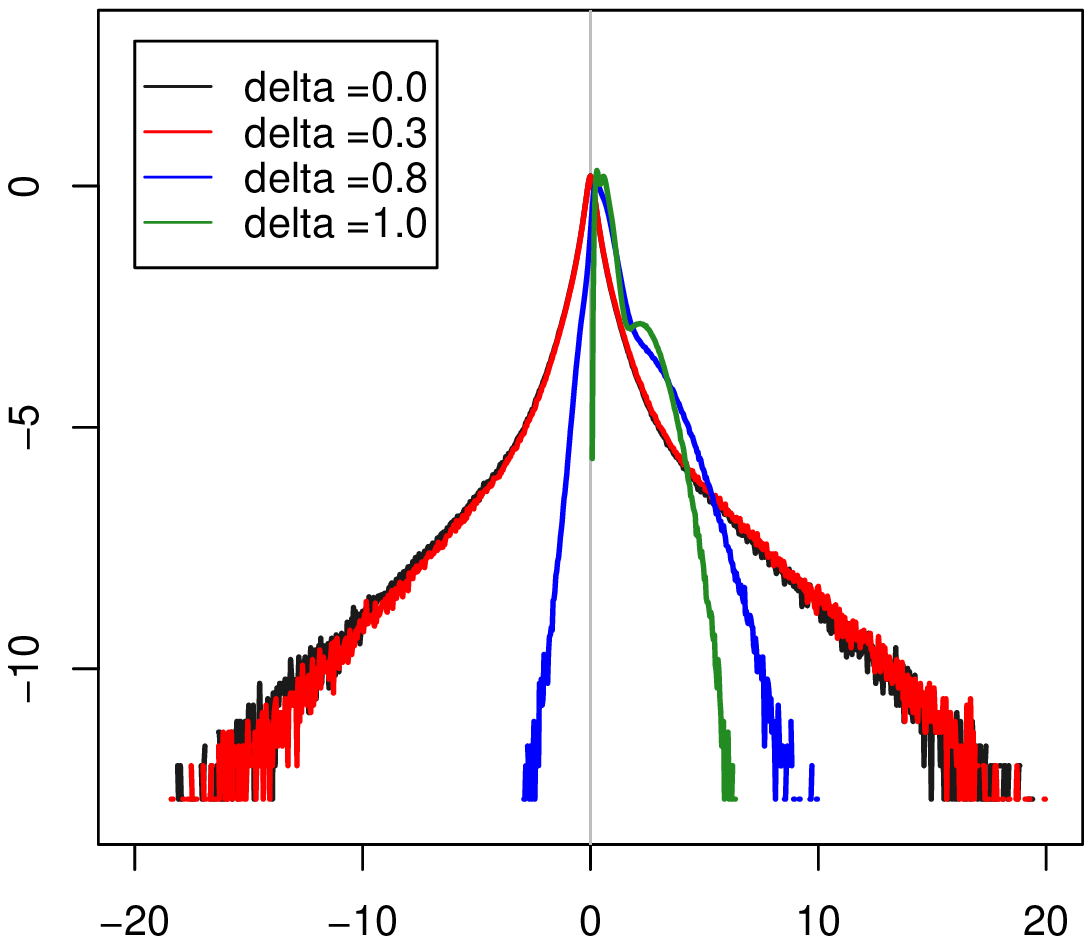}\vspace{-12mm}\\
\hspace{10mm} $ v_i $\\
\end{minipage}
\begin{minipage}{80mm}
(b)\vspace{-10mm}\\
\rotatebox{90}{\hspace{35mm}\rotatebox{0}{$ \log d(v_i) $}}\hspace{-4mm}
\includegraphics[width=7cm]{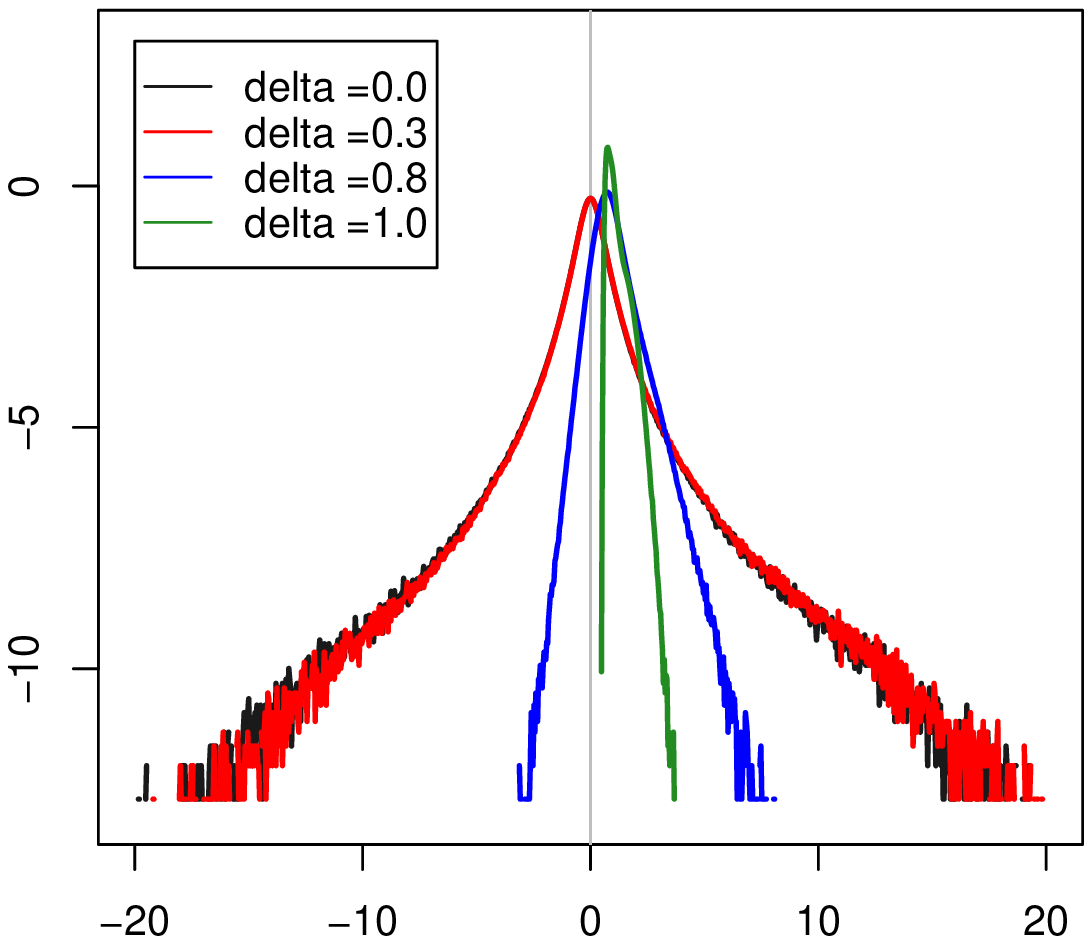}\vspace{-12mm}\\
\hspace{10mm} $ v_i $\\
\end{minipage}
\caption{Densities of the first eigenvector $ \bmv $ for a model.  (a) is the case that network nodes are degrees 4 and 12 with a ratio of 0.9:0.1. (b) is degrees 4 and 6.  The sizes of matrices are $ N=2^{12} $.}
\label{fig2d4-12-09eigenvector}
\end{figure}

\begin{figure}
\begin{minipage}{80mm}
\rotatebox{90}{\hspace{35mm}\rotatebox{0}{$ \log d(v_i) $}}\hspace{-4mm}
\includegraphics[width=7cm]{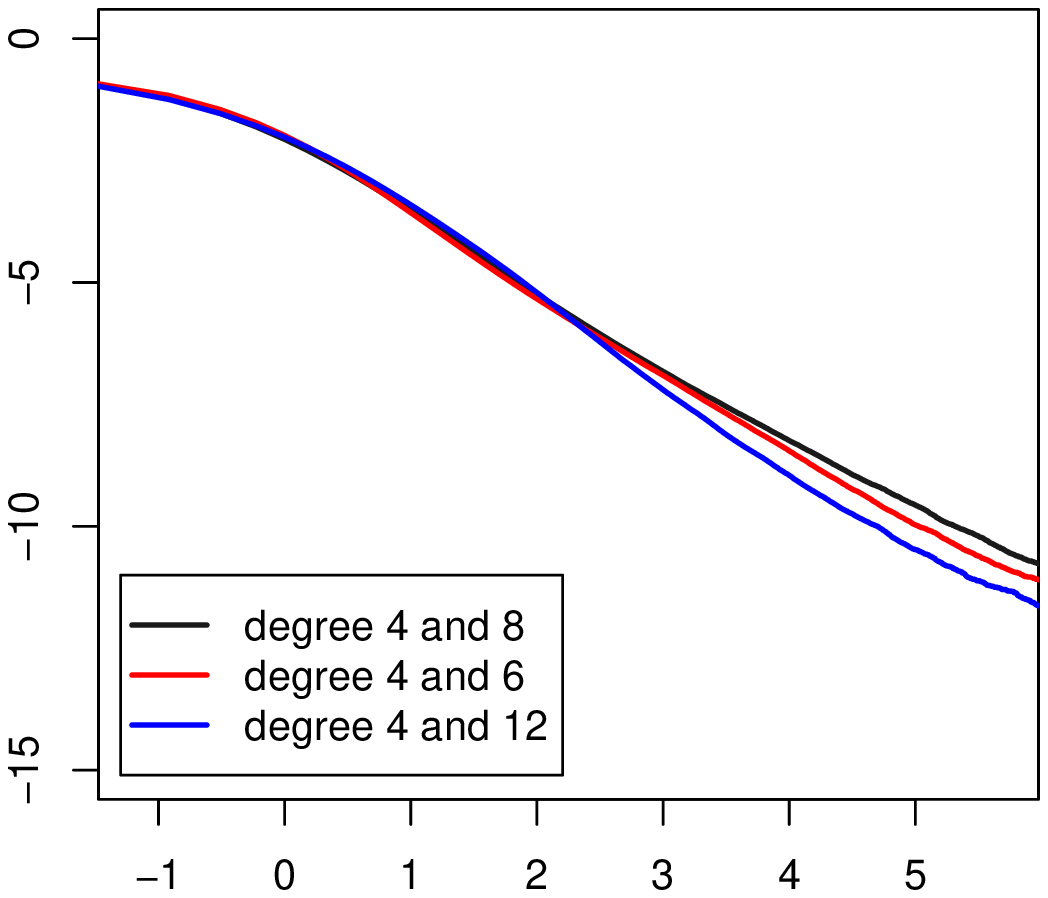}\vspace{-12mm}\\
\hspace{10mm} $ \log v_i $\\
\end{minipage}
\caption{Cumulative distribution of $ v_i $ for the cavity method.  Network nodes  constructed by degrees 4 and 6, 4 and 8, 4 and 12 are compared.  $ \varDelta = 0.0 $.}
\label{fig2d4tails}
\end{figure}

In Fig.~\ref{fig2d4-12-09eigenvector}(a), we shows the result of the density functions $ d(v_i) $ for the model such that the network has two different types of degrees for its nodes and the degrees are 4 and 12 with a ratio of 0.9:0.1.  In Fig.~\ref{fig2d4-12-09eigenvector}(b), the degrees are 4 and 6.  These figures are similar to the case of the degrees 4 and 8 with the ratio of 0.9:0.1, see Fig.~\ref{fig4-8-09eigenvector}.  In Fig.~\ref{fig2d4tails}, we shows the cumulative distribution of $ v_i $ for the network whose larger degree of nodes is 6, 8 and 12 when $ \varDelta = 0.0 $. We find those are similar to each other.

Here, we conclude the followings for 2-DTD network:
\begin{enumerate}
\item The cavity method works sufficiently well for the 2-DTD model.
\item The critical points evaluated by the scaling method and estimated by the cavity method are not in agreement, although those are close.  The finite size scaling correction might be necessary.
\item Concerning $ \varDelta < \varDelta_c $, the density function $ d(v_i) $ decays with the power law if the ratio of the nodes whose degree is larger is sufficiently small, although the density function $ d(v_i) $ decays exponentially when the network has sufficiently many nodes whose degree is larger. 
In addition, the magnitude of the largest degree does not affect the slope in the tail of the density function significantly. 
\item Concerning $ \varDelta > \varDelta_c $, the density function $ d(v_i) $ decays exponentially in any case.
\item The larger values of $ v_i $ originate from the nodes whose degree is larger, and the smaller values of $ v_i $ originate from the nodes whose degree is smaller, and vice versa.
\end{enumerate}


\subsection{Poissonian network model}
\begin{figure}
\rotatebox{90}{\hspace{35mm}\rotatebox{0}{$\log\, \varLambda $}}
\includegraphics[width=7cm]{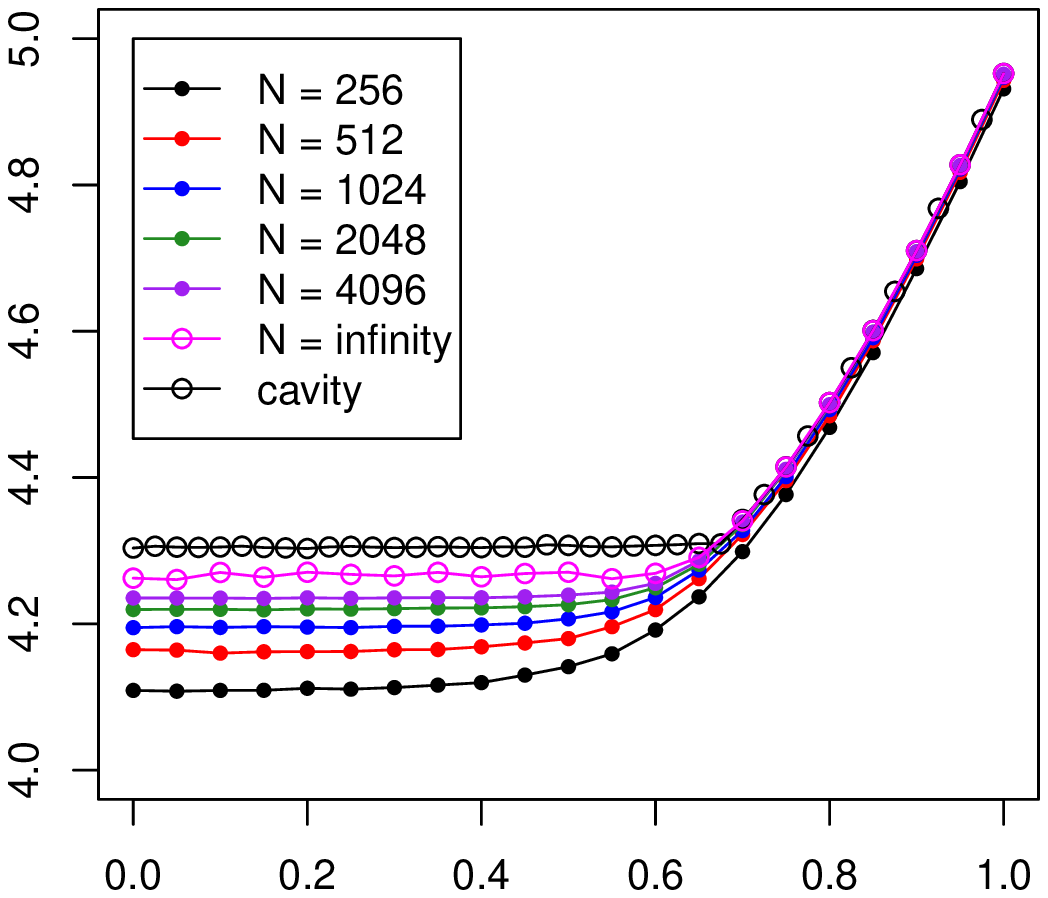}\vspace{-10mm}\\
\hspace{10mm} $ \varDelta$
\caption{First eigenvalue $ \varLambda $ versus $ \varDelta $ of networks those degree distribution of nodes is the modified Poisson distribution, under the condition that the maximum degree is 8.}
\label{figure4-8Poisson-Lambda}
\end{figure}

Figure \ref{figure4-8Poisson-Lambda} shows the results of the first eigenvalues of the Poissonian network evaluated both by the cavity and power methods to the adjacency matrices.  To eliminate the monopoly effect from very large degree vertices, we restrict the largest degree of this network as 8 ($ =k_{\rm max}$).  For this reason, our definition of the Poissonian model is as Eq. (\ref{normalize-poissonian}), as mentioned.  To obtain these results in Fig.~\ref{figure4-8Poisson-Lambda}(a), we take the average of 2000 configurations.  The eigenvalues $ \varLambda $ increase with the system size $ N $ which is the same as for 2-DTD model.  In addition, the results from the cavity method are slightly larger than those from the power method with the scaling method, especially at the lower region of $ \varDelta $.  In this case, the difference is larger than the case of 2-DTD network.  This result might originate from the increasing complexity of the network structure.

\begin{figure}
\rotatebox{90}{\hspace{35mm}\rotatebox{0}{$ \beta $}}
\includegraphics[width=7cm]{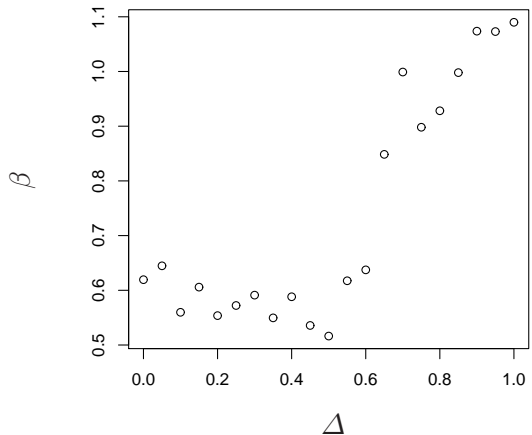}\vspace{-10mm}\\
\hspace{10mm} $ \varDelta $
\caption{Values of $ \beta $.}
\label{fig-Poisson-beta-delta}
\end{figure}

Assuming Eq.~(\ref{assumptionLambda(N)=..}), we find the values of $ \varLambda $ for $ N=\infty $ as in the previous section.  The data fit as well as those of 2-DTD model.  Figure \ref{fig-Poisson-beta-delta} shows the values of $ \beta $ for the Poissonian network model.  These results are similar to those of 2-DTD model.  In Fig.~\ref{figure4-8Poisson-Lambda}, the values of $ \beta $ are around 0.6 in the region that the line of $ \varLambda $ evaluated by cavity method becomes flat.  In other hand, in the region that the value of $ \varLambda $ increase with the value of $\varDelta$, the values of $ \beta $ are around 1.0. These results are similar to those of 2-DTD network model.  From above all, the Poissonian network model might have the critical point between the areas in which the line of $ \varLambda $ is flat and has a slope.

\begin{figure}
\begin{minipage}{80mm}
(a)$\varDelta = 0.0$\vspace{-10mm}\\
\rotatebox{90}{\hspace{30mm}\rotatebox{0}{$ \log d(v_i) $}}\hspace{-5mm}
\includegraphics[width=7cm]{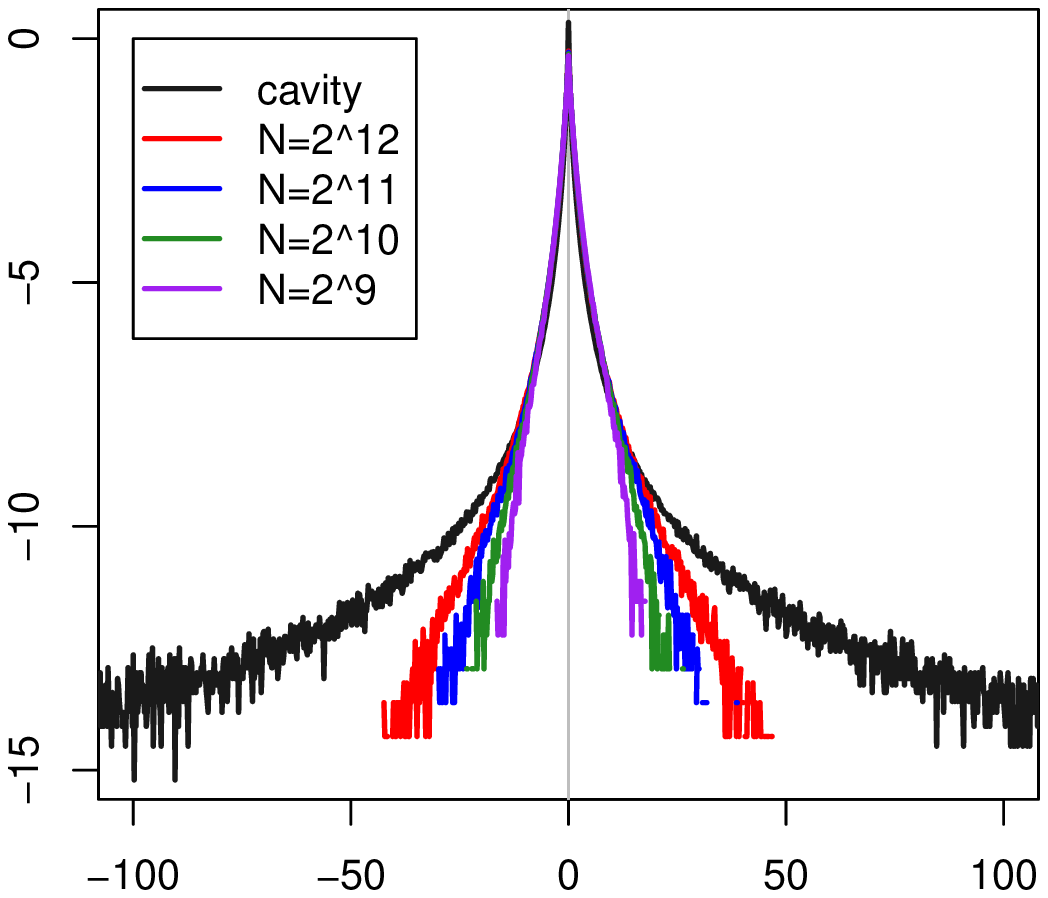}\vspace{-12mm}\\
\hspace{10mm} $ v_i $ \vspace{-62mm}\\
\hspace{35mm}\includegraphics[width=3cm]{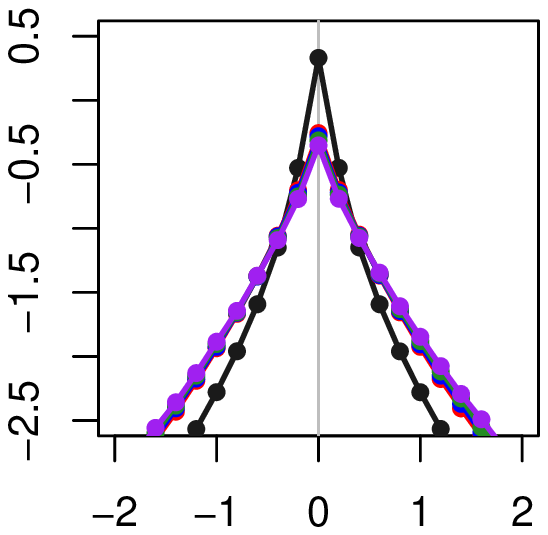}\vspace{-12mm}\vspace{39mm}\\
\end{minipage}
\begin{minipage}{80mm}
(b)$\varDelta = 0.8$\vspace{-10mm}\\
\rotatebox{90}{\hspace{30mm}\rotatebox{0}{$ \log d(v_i) $}}\hspace{-5mm}
\includegraphics[width=7cm]{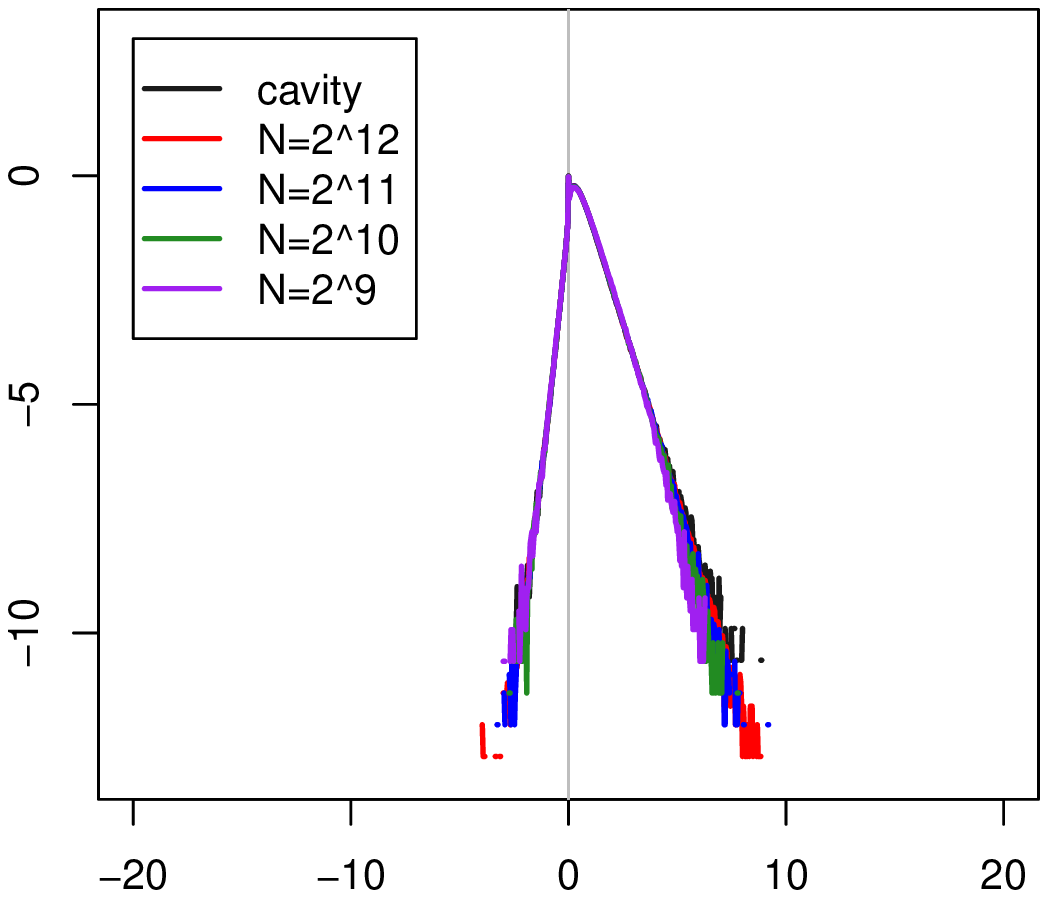}\vspace{-12mm}\\
\hspace{10mm} $ v_i $\vspace{-62mm}\\
\hspace{35mm}\includegraphics[width=3cm]{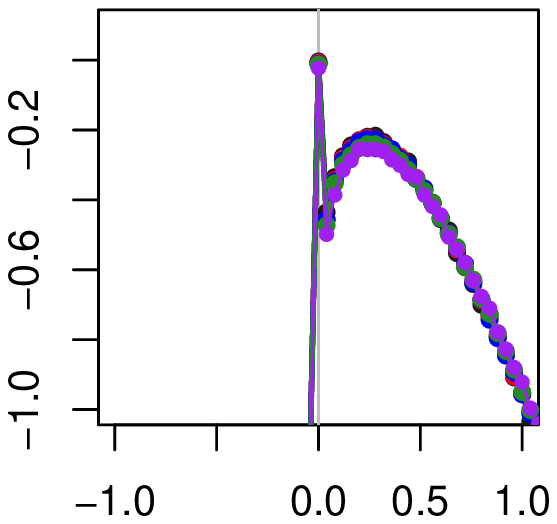}\vspace{-12mm}\vspace{39mm}\\
\end{minipage}
\vspace{5mm}

\begin{minipage}{80mm}
(c)$\varDelta = 0.0$\vspace{-10mm}\\
\rotatebox{90}{\hspace{30mm}\rotatebox{0}{$ \log d(v_i) $}}\hspace{-5mm}
\includegraphics[width=7cm]{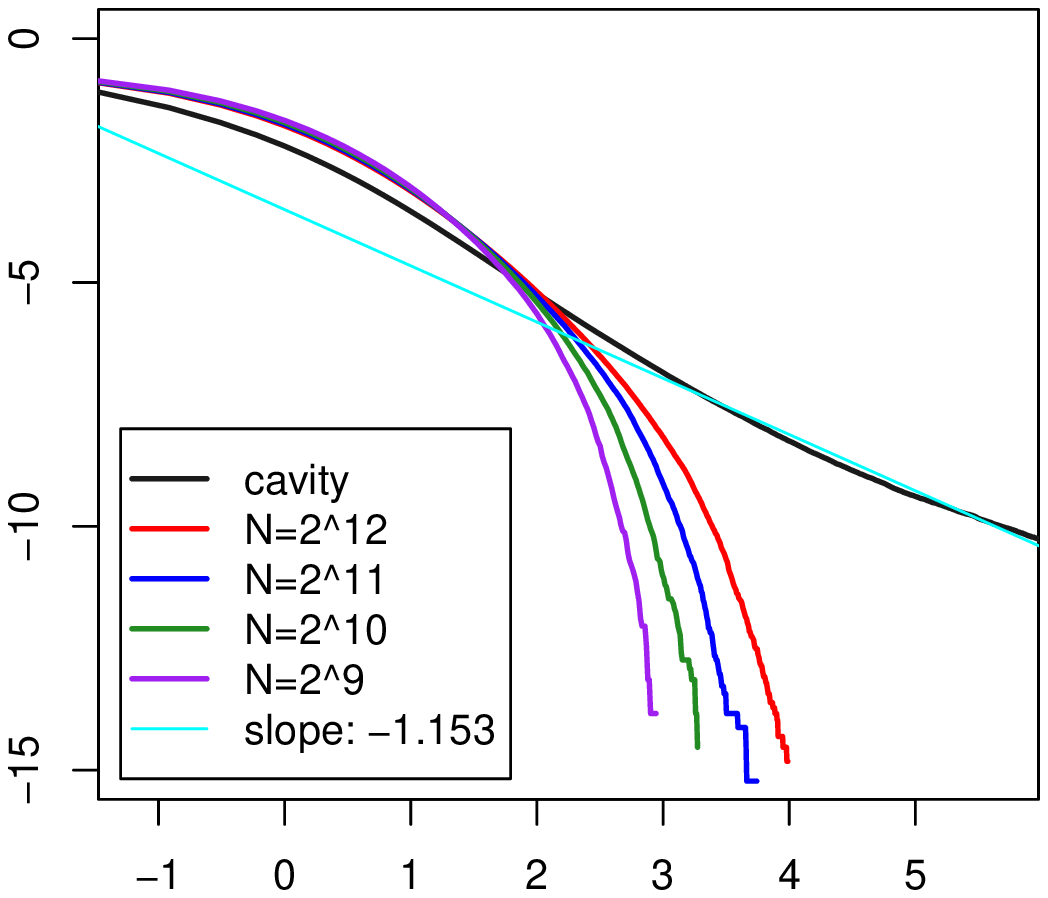}\vspace{-12mm}\\
\hspace{10mm} $ \log v_i $\\
\end{minipage}
\begin{minipage}{80mm}
(d) Convergence\vspace{-10mm}\\
\rotatebox{90}{\hspace{30mm}\rotatebox{0}{$ \log S $}}\hspace{-5mm}
\includegraphics[width=7cm]{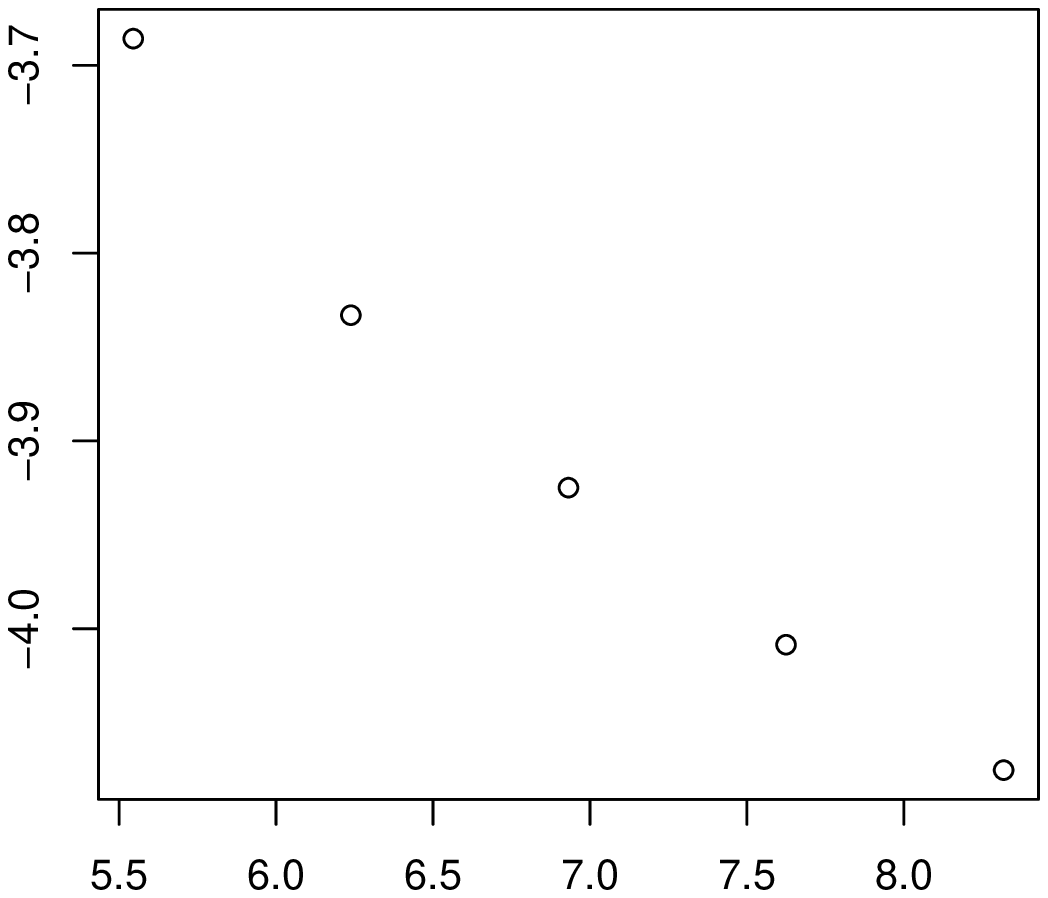}\vspace{-12mm}\\
\hspace{5mm} $ \log N $\\
\end{minipage}
\caption{Density of the first eigenvector $ \bmv $ for the Poissonian network.  (a) is for $\varDelta = 0.0$ and it is normalized as $ \sum_i |v_i| = N $.  The inset is its magnification around the region that is close to the origin with respect to $ v_i $-axis.  (b) is for $\varDelta = 0.8$ and normalized as $ \sum_i v_i^2 = N $.  The inset is its magnification around the region that is close to the origin with respect to $ v_i $-axis.  (c) is the log-log plot of the cumulative distribution of (a) for the region where $v_i$ is positive.  (d) shows convergence to the result of the cavity method evaluated by Eq.~(\ref{convergence-assess}) when $ N \to \infty $.}
\label{fig4-8Piosson-eigenvector}
\end{figure}

Figures \ref{fig4-8Piosson-eigenvector}(a) and (b) show the density function $ d(v_i) $ of the first eigenvectors $ \bmv $ based on the power method and the values $ v_i\ (= H_i/A_i) $ based on the cavity method on the Poissonian network model.  In Fig.~(a), $ \varDelta =0.0 $ and in (b), $ \varDelta =0.8 $.  In each figure, we compare the density functions for different system sizes $ N $ and the results obtained by the cavity method.  Figure \ref{fig4-8Piosson-eigenvector}(c) is the log-log plot of the cumulative distribution of $d(v_i)$ where $v_i$ is in the positive region.  Figure \ref{fig4-8-09eigenvector-scale}(d) shows the results of Eq.~(\ref{convergence-assess}).  We could confirm that the results obtained by the adjacency matrix converge to those obtained by the cavity method when $ N \to \infty$, calculating $S$ of Eq.~(ref{convergence-assess}).

From the above facts, we make the following conclusions:
\begin{enumerate}
\item The cavity method also works sufficiently well for the Poissonian network model.
\item Concerning the Poissonian network model, many important features such as the heavy tail for the density function and the shape of the first eigenvalue are similar to the results of the network that has only two different types of nodes with respect to the degree.  
\end{enumerate}

\section{Concluding remarks}\label{Concluding remarks}

We showed the properties of the first eigenvalue/vector for several kinds of adjacency matrices.  The cavity method is available for all those matrices.  It is often necessary to know the properties of an infinite large system in many studies such as the combinatorial problems, the random matrix problems, the network science and so on.  Therefore it is important to establish a searching method such as the cavity method to explore the properties of the infinite large system. In this paper, we mainly focused on 2-DTD model and the Poissonian model.  We also explored the Laplacian matrix and confirmed that the cavity method can be useful for that, see appendix.  

The value of the critical point of $\varDelta_{\rm c} $ for degree 4 network is still unknown.  It might not be easy to find the critical value using the methods in this paper. Therefore, it is required to find another method capable of rigorously determining the critical point $\varDelta_{\rm c} $.  The finite scaling method might be useful to determine the critical point $\varDelta_{\rm c} $, whereas it is known to be hard sometimes\cite{Takahashi}.

We showed in 2-DTD model that the density function of the eigenvectors has fat tails, i.e., it decays as the power law, when the ratio of the larger degree nodes is small; whereas the density function decays exponentially, when the majority or none of nodes are the larger degree.  
It seems that the power low decay of the density function originate from the construction of network, not from the variety of the degrees of nodes, if we consider the results of 2-DTD model and the Poissonian network model.

There must be certain rule between the degree distribution and observables.  For example, there must be a rule on the ratio of degrees so that the tail of the density becomes the power law for small $ \varDelta $ in 2-DTD model. So far, we confirmed that using the cavity method, we observe a power law decay at the tail of the density when the network nodes have two different types of the degrees which are 4 and 8 with the ratio 0.5:0.5., whereas we cannot observe a power law decay using the analysis of the adjacency matrices whose sizes are $ N=2^{12}$.  Deciding boundary of $ \varDelta_{\rm c} $ and finding the relationship between the exponent of the density's decay and the ratio of degrees will be a future study.  The relationship must also depend on the number of the larger degree if the number of the smaller degree fixed to 4.
This problem is related to the mathematical problem such that which type of transition matrix generates the stable distribution.

The reason of the similarity of the graphs of $ \beta $ in Eq.~(\ref{assumptionLambda(N)=..}), i.e.  Fig.~\ref{fig4-8-09beta-delta} and \ref{fig-Poisson-beta-delta}, seems to originate from the similarity of the properties of those two networks, because the graphs of $ \beta$ whose network nodes is only one type is not similar to those two graphs.  Here, we should remind that there are many factors deciding the scaling law between the eigenvalue $ \varLambda $ and the system size $ N $.

We only focused on the statistical values of observables, i.e. all the results in this paper are the configurational averages of observables.  It is important to study on each network that has a certain configuration.  It is also important to study on the relation between the centrality and the distance among nodes whose degree is the largest in the network.

\begin{figure}
\begin{minipage}{80mm}
$\varDelta = 0.0 $ \vspace{-8mm}\\
\rotatebox{90}{\hspace{25mm}\rotatebox{0}{$ \log d(A) $}}\hspace{-5mm}
\includegraphics[width=6cm]{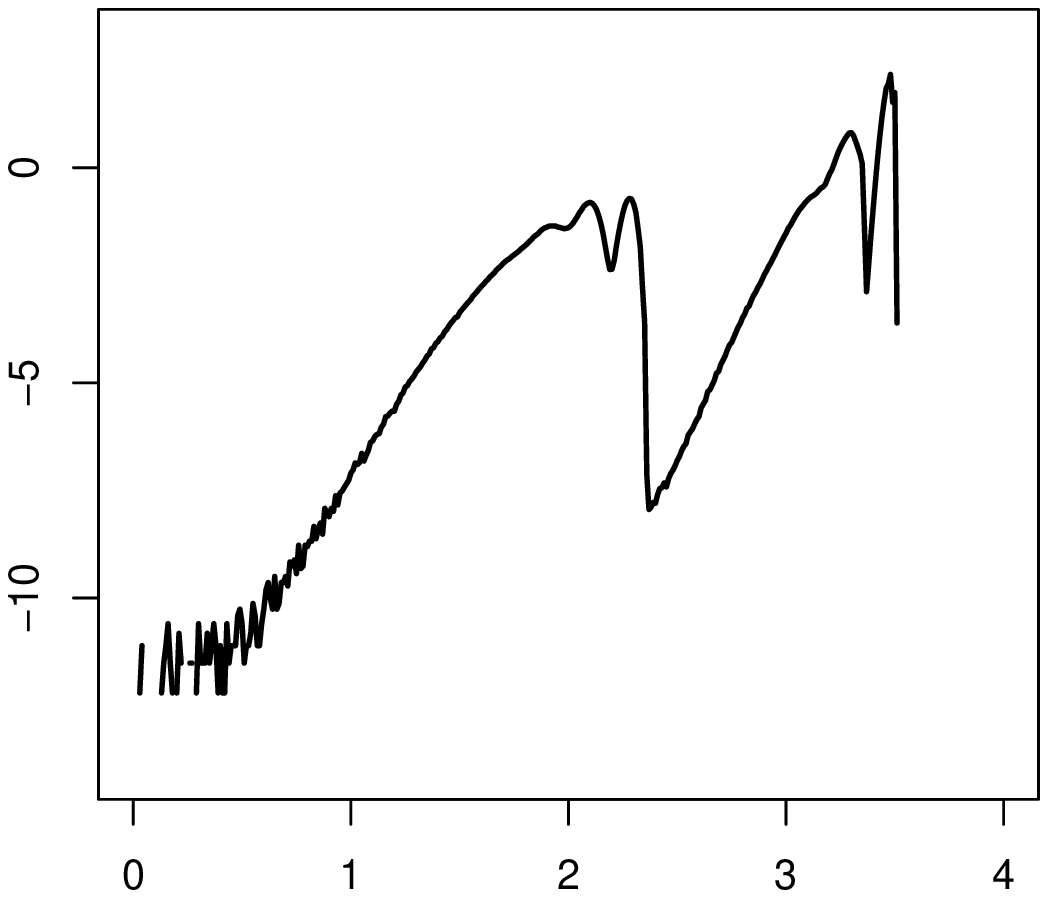}\vspace{-10mm}\\
\hspace{4mm} $ A $
\end{minipage}
\begin{minipage}{80mm}
$\varDelta = 0.8 $ \vspace{-8mm}\\
\rotatebox{90}{\hspace{25mm}\rotatebox{0}{$ \log d(A) $}}\hspace{-5mm}
\includegraphics[width=6cm]{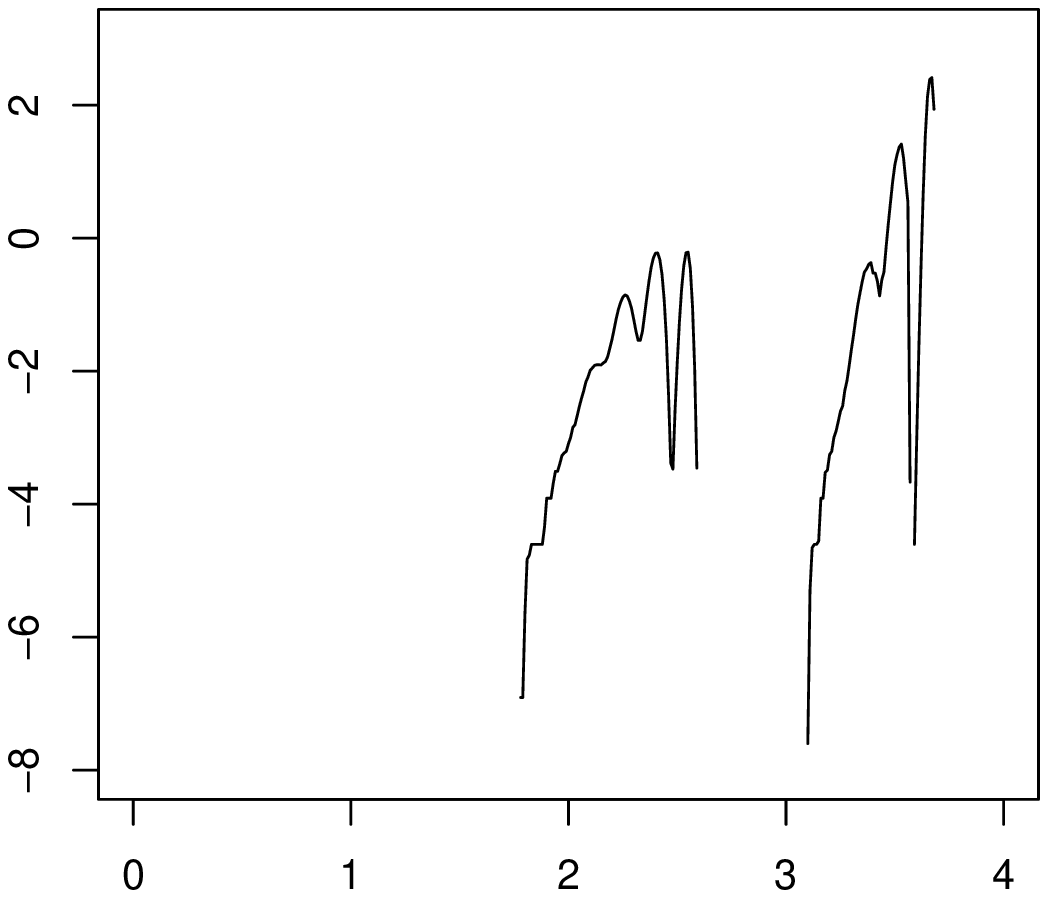}\vspace{-10mm}\\
\hspace{4mm} $ A $
\end{minipage}
\caption{Density function of the cavity field $A$ for 2-DTD network. Degrees is 4 and 8 with a ratio of 0.9:0.1, and $ \lambda \simeq \varLambda $.}
\label{fig-densA}
\end{figure}

We showed the density function of the cavity field $A$ for the 2-DTD network in Fig.~\ref{fig-densA}.  The degrees are 4 and 8 with a ratio of 0.9:0.1 under the condition that $ \lambda \simeq \varLambda $.  We found that the cavity field becomes united when $\varDelta = 0.0 $, whereas separated when $\varDelta = 0.8 $.

\clearpage
\newpage

\section*{Acknowledgements}
Authors thank to Prof. O. Watanabe and Prof. Y. Kabashima (Tokyo Institute of Technology) for their advice and help.  
We also thank to Prof. H. Maruyama and Prof. K. Minani (the Institute of Statistical Mathematics) for his help and kindness.


\appendix*
\section{Laplacian matrix}
On the basis of the adjacency matrix $\bmJ=(J_{ij})$, the Laplacian matrix $\bmJ^{({\rm L})}=(J^{({\rm L})}_{ij})$ can be made as follows:
\begin{eqnarray}
\label{def.Laplacian} 
	J^{({\rm L})}_{ij} =  - J_{ij} + \delta_{ij}\sum_j J_{ij}, 
\end{eqnarray}
where all non-zero elements of $\bmJ=(J_{ij})$ equal to 1, i.e., $ \varDelta =1$, in order to correspond with the definition of the usual Laplacian matrix.

In order to evaluate the cavity fields for the Laplacian matrix, we adjust eq.~(\ref{cavity2}) and (\ref{cavity1}) as follows:
\begin{eqnarray} 
&&A_{i \to l} = \lambda - J^{\rm (L)}_{ii} - \sum_{j \in \partial i \backslash l} \frac{(J^{\rm (L)}_{ij})^2}{A_{j \to i}}, 
\label{cavityLaplacian2} \\ && H_{i \to l} = \sum_{j \in \partial i \backslash l} \frac{J^{\rm (L)}_{ij} H_{j \to i}}{A_{j \to i}},
\label{cavityLaplacian1} 
\end{eqnarray}
where
$ J^{\rm (L)}_{ii} = -\sum_{j\ne i} J^{\rm (L)}_{ij} = \sum_j J_{ij}$.

From the above equations, the distribution of the cavity fields, which correspond to eq. (\ref{cavity_dist_update}), can be described as
\begin{eqnarray}
	&&
	\nonumber
	q(A,H)
	\\
	\nonumber
	&=&
	\sum_{k=1}^{k_{\rm max}} \! r(k) \! \int \! \prod_{j=1}^{k-1}  dA_j \,dH_j\, q(A_j,H_j)
	\\
	&& \hspace{10mm}
	\left\langle 
		\delta \left(A-\lambda+{\cal J}_{ii}(k)+\sum_{j=1}^{k-1}\frac{{\cal J}_j{}^2}{A_j} \right)
		\delta\left(H+\sum_{j=1}^{k-1} \frac{{\cal J}_j\, H_j}{A_j} \right ) 
	\right\rangle_{\cbJ}\! .
\label{cavity_dist_update_Laplacian}
\end{eqnarray}
where 
\begin{eqnarray}
	{\cal J}_{ii}(k) = \sum_{j=0}^k {\cal J}_j.
\end{eqnarray}
And the equation corresponding with eq. (\ref{fulldist}) is
\begin{eqnarray}
	&&
	Q(A,H)
	\nonumber
	\\
	\nonumber
	&=&
	\sum_{k=0}^{k_{\rm max}} \! p(k) \! \int \! \prod_{j=1}^k dA_j\,dH_j\, q(A_j,H_j)
	\\
	&&\hspace{10mm}
	\left \langle 
		\delta \left(A-\lambda+{\cal J}_{ii}(k)+\sum_{j=1}^{k}\frac{{\cal J}_j{}^2}{A_j} \right) 
		\delta \left(H+\sum_{j=1}^{k} \frac{{\cal J}_j \, H_j}{A_j} \right ) 
	\right\rangle_{\cbJ} \!.
\label{fulldist_Laplacian}
\end{eqnarray}
To calculate the equations (\ref{cavity_dist_update_Laplacian}) and (\ref{fulldist_Laplacian}), we apply the population dynamical method.

\begin{table}
\begin{tabular}{ccccccc}
\hline
\hline
$ N $ & $ 2^8 $ & $ 2^9 $ & $ 2^{10} $ & $ 2^{11} $ & $ 2^{12} $ & Cavity \\
\hline
$ \varLambda $ & 11.07 & 11.22 & 11.32 & 11.40 & 11.46 & 11.53 \\
\hline
\end{tabular}
%
\caption{First eigenvalue $ \varLambda $ for the Laplacian matrix. The networks are constructed by nodes of degrees 4 and 8 with the ratio of 0.9:0.1.}
\label{table-Lambda-Laplacian2d}
\end{table}

\begin{figure}
\begin{minipage}{80mm}
(a)$\varDelta = 1.0$\vspace{-10mm}\\
\rotatebox{90}{\hspace{35mm}\rotatebox{0}{$ \log d(v_i) $}}
\includegraphics[width=7cm]{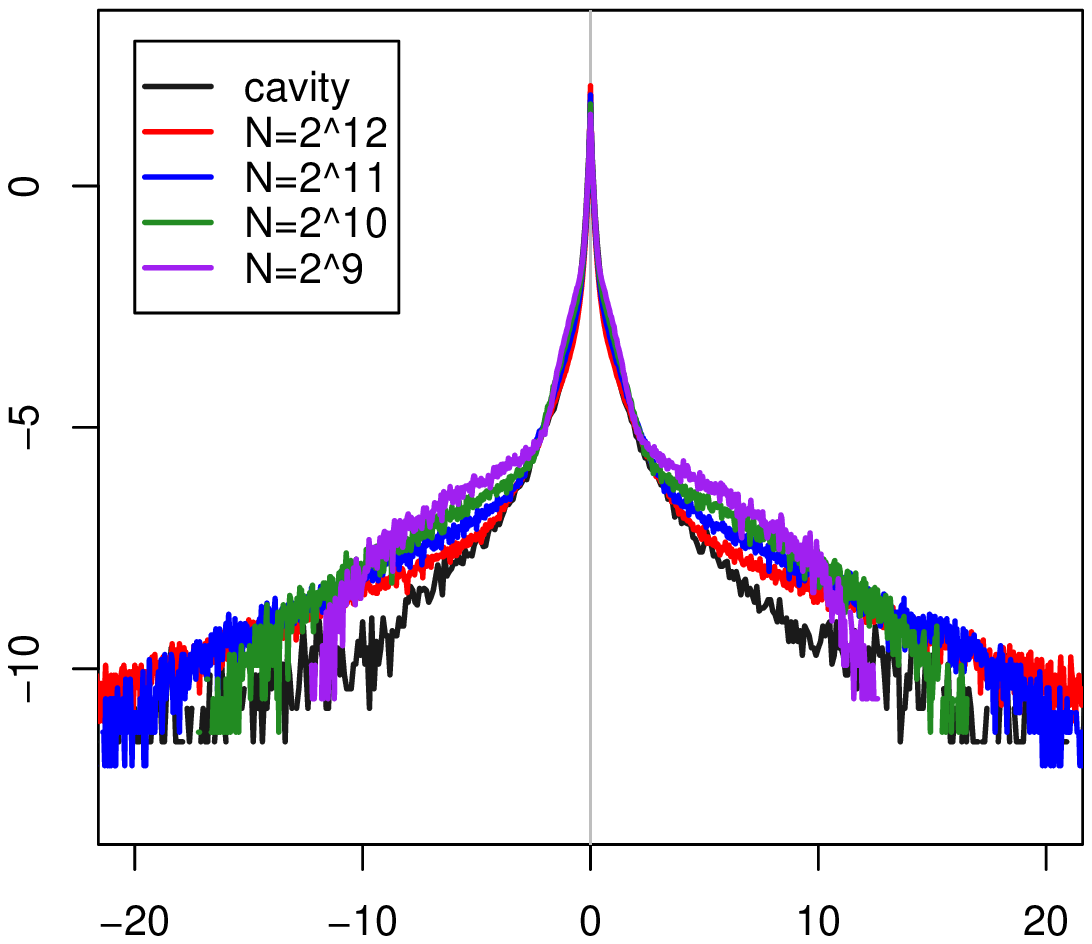}\vspace{-10mm}\\
\hspace{10mm} $ v_i $
\end{minipage}
\begin{minipage}{80mm}
(b)$\varDelta = 1.0$\vspace{-10mm}\\
\rotatebox{90}{\hspace{35mm}\rotatebox{0}{$ \log d(v_i) $}}
\includegraphics[width=7cm]{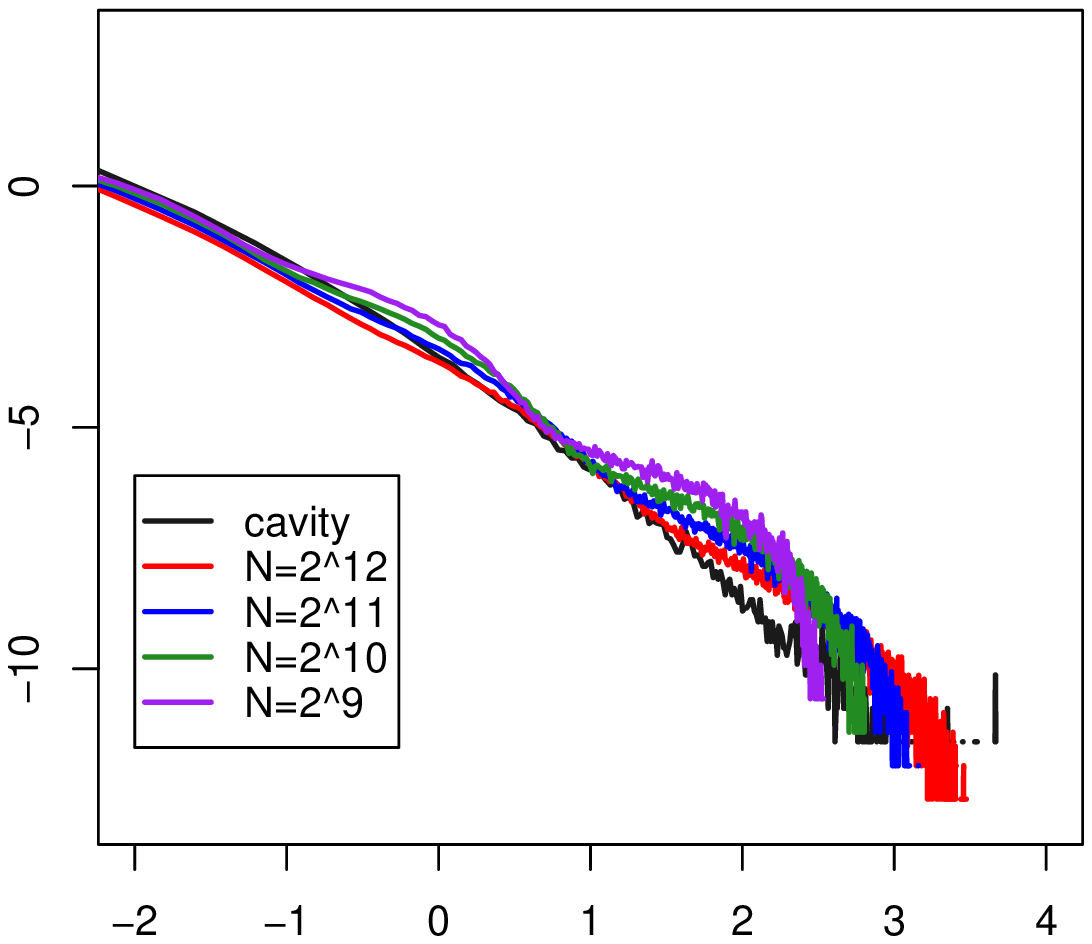}\vspace{-10mm}\\
\hspace{10mm} $ \log v_i $
\end{minipage}
\caption{Density of the first eigenvector $ \bmv $ for the Laplacain matrix derived from 2-DTD network.  (a) is the semi-log plot, (b) is the log-log plot.}
\label{fig-Laplacian2d-eigenvector-scale}
\end{figure}

Table \ref{table-Lambda-Laplacian2d} shows the eigenvalues of the Laplacian matrix that the $\bmJ=(J_{ij})$ is correspond with 2-DTD model whose degrees are 4 and 8 with a ratio of 0.9:0.1.  We show the values for several system sizes and a result of the cavity method.  It seems that the results show the convergence from small system size to the large and to the result of the cavity method.  Figure \ref{fig-Laplacian2d-eigenvector-scale} shows the density functions of the first eigenvectors $ \bmv $ and the values $ v_i\ (= H_i/A_i) $ in the cavity method.  Here we can also confirm the convergence.  

From these results, we conclude that the cavity method works sufficiently well for the Laplacian matrix model.  We also confirmed that the cavity method works sufficiently well for the Laplacian matrix model which derive from the Poissonian network model, although we does not present these results in this paper.

\clearpage

\end{document}